\RequirePackage{fix-cm}
\documentclass[natbib,smallextended]{svjour3}                     
\smartqed  
\usepackage{graphicx}
\usepackage[colorlinks=true,citecolor=blue,urlcolor=blue,breaklinks]{hyperref}
\usepackage{amsmath,amssymb}

\usepackage{mathptmx}      

\usepackage{longtable}
\usepackage{aas_macros}
\begin{document}

\title{Signatures of convection in the atmospheres of cool evolved stars}




\author{Andrea Chiavassa         \and
        Kateryna Kravchenko \and 
        Jared A. Goldberg
}


\institute{A. Chiavassa \at
              Universit\'e C\^ote d'Azur, Observatoire de la C\^ote d'Azur, CNRS, Lagrange, CS 34229, Nice,  France  \\
              and\\
   Max-Planck-Institut f\"{u}r Astrophysik, Karl-Schwarzschild-Stra\ss{}e 1, 85741 Garching, Germany\\
              \email{andrea.chiavassa@oca.eu}           
           \and
           K. Kravchenko \at
              Max Planck Institute for Extraterrestrial Physics, Gie\ss{}enbachstra\ss{}e 1, 85748 Garching, Germany              
            \and
           J. A. Goldberg \at
              Center for Computational Astrophysics, Flatiron Institute, 162 5th Avenue, 10010, New York, NY, USA 
}

\date{Received: date / Accepted: date}

\maketitle

\begin{abstract}

Evolved cool stars of various masses are major cosmic engines, delivering substantial mechanical and radiative feedback to the interstellar medium through strong stellar winds and supernova ejecta. These stars play a pivotal role in enriching the interstellar medium with vital chemical elements that constitute the essential building blocks for the formation of subsequent generations of stars, planets, and potentially even life.\\
Within the complex tapestry of processes occurring in the atmospheres of these cool and luminous stars, convection takes center stage. Convection is a non-local, complex phenomenon marked by non-linear interactions across diverse length scales within a multi-dimensional framework. For these particular stars, characterized by their considerable luminosities and extensive scale heights, convection transitions to a global scale. This transition is facilitated by the transmission of radiative energy through the non-uniform outer layers of their atmospheres.\\
To have a full understanding of this phenomenon, the application of global comprehensive 3D radiation-hydrodynamics simulations of stellar convection is of paramount importance. We present two state-of-the-art numerical codes: CO5BOLD and Athena++. Furthermore, we provide a view on their applications as: pivotal roles in enabling a comprehensive investigation into the dynamic processes linked to convection; and critical tools for accurately modeling the emissions produced during shock breakouts in Type II-P Supernovae.

\end{abstract}

\section{Introduction}
\label{intro}

Evolved cool stars of various masses are major cosmic engines, providing strong mechanical and radiative feedback on their host environment \citep{2012ARA&A..50..107L}. Through strong stellar winds and supernova ejecta, they enrich the interstellar medium with chemical elements, which are the building blocks of planets and life. In particular, these objects are known to propel strong stellar winds at speeds varying with stellar brightness, evolutionary phase, and chemical composition. Yet, stellar evolution models are not able to reproduce these winds without ''ad hoc'' physics. Therefore, a full understanding of stellar evolution in the near and distant Universe, and its impact on the cosmic environment, cannot be achieved without a detailed knowledge of wind physics across the life cycle of these stars. This requires tracing the total mass ejected as well as its composition, the velocity of the winds, and the overall geometry of the circumstellar envelope.
Evolved cool stars are objects that have reached the late phases of their evolution when the nuclear fuel in the interior is almost used up. These stars grow dramatically in size and become Asymptotic Giant Branch (AGB, the initial mass of $0.8 < M < 8\, M_\odot$, the exact value of the upper limit depends on the treatment of convection) or Red Supergiant (RSG, $M > 8\, M_\odot$) stars. AGBs are characterized by a very low effective temperature (lower than 3000 K) and a high luminosity (about 100--1000 $L_\odot$) which correspond to stellar radii of several hundred solar radii. Their high rate of mass-loss rate \citep[$\dot{M}=10^{-6}$--$10^{-4}\, M_\odot/\mathrm{yr}$, ][]{1992ApJ...397..552W} are the results of a wind acceleration mechanism based on radiation pressure on microscopic solid-state particles (dust grains), formed in atmospheres that are levitated by pulsation. State-of-the-art models are still insufficient to provide a full and consistent theory of their mass-loss \citep{2018A&ARv..26....1H}. RSGs are precursors of core-collapse supernovae with high luminosities ($L > 1000\, L_\odot$) with effective temperatures between 3450 and 4100 K and stellar radii up to several hundreds of $R_\odot$, or even more than $1000\, R_\odot$ \citep{2005ApJ...628..973L}. Several mechanisms triggering mass-loss have been discussed, including magneto-hydrodynamic waves (e.g. \citealt{2011ApJ...741...54C}), turbulent pressure from trans-sonic convection (e.g. \citealt{2021A&A...646A.180K}), and radiation pressure on molecules and dust \citep{2007A&A...469..671J}, but still there is no realistic quantitative wind model \citep{Meynet2015} that can explain the observed broad range of mass-loss rates \citep[$\dot{M}=10^{-7}$--$10^{-4}\, M_\odot/\mathrm{yr}$, ][]{2010A&A...523A..18D}. 

A full picture of all the physical processes that simultaneously trigger and shape the convective envelopes and strong winds of evolved cool stars is still missing. Indeed, the mass-loss mechanism is hard to discern because it involves a range of interacting, time-dependent physical processes on microscopic and macroscopic scales coupled with dynamical phenomenon such as convection and pulsation in sub-photospheric layers, strong radiating shocks in the atmosphere, and dust condensation as well as radiative acceleration in the wind forming regions \citep{2018A&ARv..26....1H}. By depositing angular momentum, (sub)stellar companions are known to shape the outflow of cool evolved stars \citep{2020Sci...369.1497D}.


In this framework, three physical ingredients should play a predominant role, depending on the stellar type, in initiating and maintaining the high mass loss. In the first place, the evolution of these objects is impacted by stellar convection. Convection is a difficult process to understand because it is non-local, three-dimensional, and involves non-linear interactions over many disparate length scales, and it is partially responsible for transporting heat up to the visible surface \citep{2009LRSP....6....2N}. In AGB and RSG atmospheres, convection is inferred from a few giant structures observed at the stellar surface with sizes comparable to the stellar radius and evolving on weekly or yearly time scales (e.g. \citealt{2023arXiv231116885M,2021Natur.594..365M,2018Natur.553..310P,2018A&A...614A..12M,2010A&A...515A..12C,2011A&A...528A.120C}). Such large-scale convective structures are expected theoretically due to the convective plume size approximately scaling with the larger pressure scale-heights in these giant envelopes \citep[e.g.][]{BohmVitense1958, 2014MNRAS.445.4366T, 2011A&A...535A..22C, 2017A&A...600A.137F, Goldberg2022a}.
This results in more extreme atmospheric conditions than in the Sun: very large variations in velocity, density, and temperature produce strong radiative shocks in their extended atmosphere that can cause the gas to levitate and thus contribute to mass-loss \citep{2011A&A...535A..22C,2017A&A...600A.137F,2019A&A...623A.158H,2023A&A...669A..49A}. In the outer layers of these extreme atmospheres, radiation begins to carry non-negligible amounts of the flux, yet the the opacity, set by the (varied) density and temperature, can experience order-of-magnitude fluctuations from H and He opacity peaks. Because the luminosities are so high, these fluctuations in opacity will affect the dynamics of the convective envelope.\\

The second ingredient is dust. In the case of the slow and massive winds of AGB stars, the wind is commonly assumed to be driven by radiative pressure on dust grains, forming in the extended stellar atmospheres \cite{2018A&ARv..26....1H}. The pulsations, together with large-scale convective flows, trigger strong radiative shock waves in the stellar atmospheres, which intermittently lift gas to distances where temperatures are low enough to let dust form \cite{2019A&A...623A.158H}. This has been recently shown in the theoretical work of \cite{2023A&A...669A.155F}, where the simulations are capable to follow the flow of matter from the stellar interior, through the dynamical atmosphere, and into the wind-acceleration region, while taking non-equilibrium dust formation and the interaction of matter with radiation into account. 
The RSG phase may also be characterised by the presence of dust close to a star \citep{2019A&A...628A.101H,2021Natur.594..365M} as well as the inferences of circumstellar matter in the pulsating progenitor of a Type II supernova \citep{2022ApJ...934L..31K}. A similar picture as for AGB stars could also be true for RSGs where pulsation-driven  winds may dominate (e.g. \citealt{2010ApJ...717L..62Y}).


The third ingredient is the magnetic field. \cite{2011ApJ...741...54C} presented a predictive description of mass-loss, based on Alfv\'en-wave-driven wind. Alfv\'en waves (i.e., incompressible, transverse magneto-hydrodynamic waves) are expected to provide a source of heating by dissipation. It should be noted that Alfv\'en-wave-driven winds require open field lines, radially directed away from the star, in order for the gas to be accelerated and escape \citep{2018A&ARv..26....1H}. RSGs and AGBs have low-intensity integrated magnetic fields (of the order of 1-10 Gauss) that have been identified and monitored over several years \citep{2017A&A...603A.129T,2014A&A...561A..85L,2010A&A...516L...2A}. However, the knowledge of the magnetic field strength and geometry is still limited and its origin from a possible astrophysical dynamo in these stars would most likely be very different from the dynamo at work in solar-type stars due to the small radiative core, slow rotation and the fact that only a few convection cells are present at their surface at any given time \citep{2010A&A...516L...2A,2002AN....323..213F}. 

To provide quantitative constraints on the physics of cool evolved stars, observational techniques have reached such a level of excellence that it is now possible to reconstruct spatially resolved images of stellar surfaces in the near IR with interferometric (e.g. \citealt{2022A&A...658A.185C,2020A&A...640A..23C,2018A&A...614A..12M,2018Natur.553..310P,2017Natur.548..310O,2009A&A...508..923H}) or astrometric techniques \citep{2022A&A...661L...1C,2018A&A...617L...1C}. Simultaneous intensity and polarization stellar surface images have been obtained at unprecedented resolutions with SPHERE@VLT (e.g. \citealt{2016A&A...585A..28K}). Images obtained with ALMA were used to measure the rotation of the molecular envelope of the RSG star Betelgeuse \citep{2018A&A...609A..67K}. Finally, spectropolarimetric observations (with the instruments Narval-TBL and ESPaDOnS-CFHT) have made it possible, among other things, to detect and characterize the surface magnetism and dynamics of RSGs \citep{2022A&A...661A..91L,2018A&A...615A.116M,2010A&A...516L...2A} and AGBs \citep{2019A&A...632A..30L,2014A&A...561A..85L}.\\


In the following Sections, we concentrate on the stellar atmospheric dynamics based on convection-related processes, leaving further reviews on dust- or magneto-related subjects for later works. 


\section{3D radiation-hydrodynamics simulations of stellar convection}\label{sec_cobold}

The interpretation of high-resolution spatially and spectrally resolved observations of evolved stars requires realistic modeling that takes into account most of the processes at work in the atmosphere (e.g., convection, shocks, pulsation, radiative transfer, ionization, molecules and dust formation, magnetic fields). 

Despite the apparent sphericity of stars, all the physical processes happening in their atmospheres are truly three-dimensional and therefore, they require a three-dimensional treatment, unlike the traditional one-dimensional approach in stellar-evolution modeling. For instance, one crucial point is the radiative transfer treatment that has to be simultaneously solved through a highly turbulent medium with large density variations over optical depths ranging from far above unity down to the photospheric regions where the radiation's contribution is larger. Another related point is the convection-related processes, that involve non-linear interactions over many disparate length scales. In this context, the use of numerical 3D radiation-hydrodynamics simulations of stellar convection is extremely important. In recent years, with increased computational power, it has been possible to compute grids of 3D radiation-hydrodynamics (RHD) simulations of the whole stellar envelope that are used to predict reliable synthetic spectra and images for several stellar types. 
The first realistic simulations of solar granulation were performed by \cite{1982A&A...107....1N} and included 3D, time-dependent hydrodynamics (with moderate spatial resolution) and non-local radiative energy transfer, already then with a simple treatment of the frequency-dependence of the opacities. After this pioneering work and more recently several 3D simulations (in "box-in-a-star" configuration) grids have been produced for a substantial portion of the Hertzsprung-Russell \citep{2013A&A...557A..26M,2013ApJ...769...18T,2009MmSAI..80..711L}. They are used to obtain precise determinations of stellar parameters  (e.g. \citealt{2011A&A...534L...3B,2012A&A...545A..17C}), radial velocity (e.g. \citealt{2013A&A...550A.103A,2018A&A...611A..11C}), chemical abundance (e.g. \citealt{2009ARA&A..47..481A,2011SoPh..268..255C,2022A&A...661A.140M}), photometric colours (e.g. \citealt{2018A&A...611A..68B,2018A&A...611A..11C}), and on planet detection/characterisation 
(e.g. \citealt{2015A&A...573A..90M,2019A&A...631A.100C,2019Geosc...9..114C}). Moreover, these grids have yielded a physical understanding of the nature of RHD convection, proving a quantitative way to set the outer boundary condition in 1D stellar models \citep{2014MNRAS.445.4366T,Salaris2015,2016A&A...592A..24M,2018MNRAS.478.5650M,Sonoi2019}.

However, the large sizes of the turbulent, extended outer envelope of evolved stars (such as AGB and RSG) manifest inherently 3D convective properties that need the use of global (i.e., including a major part of the convective envelope as well as the near environment of the star) and CPU-intensive simulations. In these stars, the convection-related structures are comparable to the stellar diameter and need a large enough domain. To account for that, global "star-in-a-box" simulations are used: they include the majority of the convective envelope as well as the near environment of the star and differ essentially in boundary conditions and the gravitational potential from local "box-in-a-star" models that use constant surface gravity (i.e. plane-parallel approximation) and simulate only a relatively small area and shallow depths of a star with parameters close to the main sequence and Red Giant Branch stars. The "star-in-a-box" simulations return very large surface contrast, due to the combination of violent convective flows and high-amplitude waves, which is very different from the solar case of both smaller convective structures and lower mode amplitudes \citep{2012JCoPh.231..919F}. There are two numerical codes that are able to perform such simulations: (i) the CO5BOLD code \citep{2012JCoPh.231..919F} has been used to simulate AGB \citep{2017A&A...600A.137F,2023A&A...669A..49A} and RSG \citep{2011A&A...535A..22C} stars; (ii) the Athena++ code \citep{2020ApJS..249....4S} that can simulate RSG stars \citep{Goldberg2022a} in addition to other luminous massive stars, \citep{2018Natur.561..498J,2022ApJ...924L..11S,Schultz2023}. With these two numerical approaches, it is possible to follow the flowing matter in complete 3D geometry: from the deep internal layers, where non-stationary convection produces sound waves; to the surface, where the otherwise standing waves can partially escape from the acoustic cavity of the interior, and speed up due to the exponential decrease in density and turn into shocks, that propagates all the way to the outer circumstellar envelope.

\subsection{CO5BOLD and Athena++ codes}


CO5BOLD solves the coupled equations of compressible hydrodynamics and non-local radiative energy transport in the presence of a fixed external spherically symmetric gravitational field on a 3D cartesian grid. It is important to include an accurate gravitational potential because the mass in the simulation domain is not negligible and the envelope is loosely bound. The hydrodynamics scheme of CO5BOLD uses a finite-volume approach. By means of operator (directional) splitting \citep{1968SJNA....5..506S}, the 2D or 3D problem is reduced to one dimension. To compute the fluxes across each cell boundary in every 1D column direction, an approximate 1D Riemann solver of Roe type \citep{1986JCoPh..63..458R} is applied, including appropriate modifications to account an adapted Equation of State, a non-equidistant grid, and a gravitational potential \citep{2012JCoPh.231..919F}. CO5BOLD can also handle the calculation of the magneto-hydrodynamic equations \citep{2022A&A...660A.115R} using a flux-conserving HLL solver \citep{harten1983}. Finally, CO5BOLD may account for rotation performing calculations in a corotating frame \citep{2017A&A...600A.137F}: a centrifugal potential is added to the gravitational potential resulting in an application of the Coriolis force at each hydrodynamics step.
All six surfaces of the computational box employ the same open boundary condition. The radiative transport is performed using a short-characteristics method \citep{2012JCoPh.231..919F} using Rosseland mean opacity tables \citep[merged at around 12\,000\,K from high-temperature OPAL data and low-temperature PHOENIX data, ][ respectively]{1992ApJ...397..717I,1997ApJ...483..390H}. Different works report global simulations employing  three or five opacity bins \citep{2011A&A...535A..22C,2019A&A...623A.158H, 2023A&A...669A.155F}.

Athena++ is an adaptive mesh refinement code solving the hydrodynamic equations based on a task-based dynamic execution \citep[similar to what has been implemented in other codes such as DISPATCH, ][]{2018MNRAS.477..624N}. To provide robust and accurate shock capturing, a Godunov-type method is used 
to ensure the non-oscillatory spatial reconstruction of the fluid variables at cell interfaces, 
a Riemann solver to compute upwind fluxes and electric fields at cell faces, 
and a time-integration algorithm to advance the solution \citep{2020ApJS..249....4S}. Athena++ uses distributed-memory parallelism through domain decomposition and is capable of employing both non-relativistic and relativistic magnetohydrodynamics. The simulations of RGS stars use a spherical polar coordinate system with 256 bins in azimuthal angle and 128 uniform bins in polar angle, assuming a non-rotating stellar model and neglecting magnetic fields \citep{Goldberg2022a}. The code includes an accurate spherically symmetric gravitational potential $G M(r)/ r$ which is a function of both the mass inside the simulation domain's inner boundary as well as the mass in the simulated envelope ($M(r)= M_{\rm IB} + \Delta m(r)$ where $\Delta m$ is the total enclosed mass at radius $r$ from summing each successive shell).
The radiative transport is fully coupled to the hydrodynamics in Athena++ directly solving the time-dependent, frequency-integrated (grey) radiative transport equation for specific intensities over discrete ordinates \citep{Jiang2021} with 120 angles. Radiation is solved for in the co-moving frame and then self-consistently accounted for as a source term in both the energy and momentum equations by transforming to the lab frame. The Rosseland and Planck mean absorption opacities are determined by interpolation of the OPAL opacity tables \citep{1996ApJ...464..943I}. Similar numerical setups have been successfully used to model both local patches and global simulations of stellar envelopes in different locations of the HR diagram \citep{2015ApJ...813...74J,2018Natur.561..498J,2022ApJ...924L..11S,Schultz2023}. The ability to handle multi-group opacities exists in Athena++ \citep{2022ApJS..263....4J}, but is presently untested for stellar envelopes.

\subsection{Brief summary of results and mutual comparison}

The first CO5BOLD simulation were published by \cite{2002AN....323..213F} and, since then, have proved to be realistic and indispensable tools for the understanding of the various complex dynamical processes in evolved stars. Since 2010, CO5BOLD simulations have been used to interpret the large majority of interferometric imaging observations \citep[for a short summary of imaging programs on stellar convection, see ][]{2018Msngr.172...24P} with the VLTI and CHARA telescopes: in fact, large convection-related structures and large intensity contrasts render these features observable in detail with current interferometers. More recently, CO5BOLD simulations have also been used to interpret convection cycles using spectroscopy (e.g. \citealt{2021A&A...650L..17K} and Section~\ref{sec_cycles}).\\
While CO5BOLD simulations have been available since \cite{2002AN....323..213F}, the employment of Athena++ to compute RSG stars is very recent and and already important results have been reported. \cite{Goldberg2022a} showed an unusual outcome of an inverse correlation between density and radial velocity in the outer layers, uncharacteristic of typical stellar convection, and a density structure much more extended than 1D stellar evolution predictions. The authors also managed to calibrate the mixing length ($\alpha$ = 3 -- 4) for the mixing length theory (MLT), showing that the radiation pressure provides $\approx$ 1/3 of the support against gravity in this region. The extended and inhomogeneous atmosphere has also an impact on the shock breakout of Type II-P supernovae, as shown by \cite{Goldberg2022b}: the large-scale order-of-magnitude 3D fluctuations in density cause the shock to break out at different radii at different times, and the duration of the shock breakout is prolonged over $\approx$3 -- 6 hr, smeared out by the shock-crossing time of the inhomegeneous outer layers. The authors conclude that the intrinsically 3D nature of the envelope precludes the possibility of using observed UV rise times to measure the stellar radius via light-travel time effects, but enables shock-breakout observations to probe the scale of the convective inhomogeneities when combined with additional information about the shock velocity from other supernova observables.\\
A quantitative comparison between the results of the two codes is reported in \cite{Goldberg2022a}, with a few points: (i) CO5BOLD code neglects the radiation pressure in deeper layers of the star (an approximation that is only acceptable for low optical depth, where radiative transport dominates and where the observed flux is formed), inhibiting the ability to correctly simulate the deeper nearly-constant-entropy convective zone there. 
As a consequence, the CO5BOLD RSG simulations exhibit a positive entropy gradient at about $\approx$75$\%$ of the stellar radius \citep{2011A&A...535A..22C} while Athena++ simulation predicts a nearly-flat, declining entropy profile ($ds/dr\lesssim 0$) throughout the convective envelope more consistent with 1D models of stellar convection. (ii) Both codes show coherent convective plumes across the star with comparable radial velocities of tens of km/s and comparable convective overturn times of hundreds of days. (iii) Simulations with both codes display high-entropy material above the expected stellar photospheric radii with density fluctuations developing far from the stellar photosphere. (iv) Both codes predict that convection contributes significantly to the energy transport in the stellar interior and (v) that the turbulent energy density from the vigorous convective motions dominates over the thermal energy in the region above the photospheric stellar radius.


\section{The dynamic atmosphere of evolved stars as seen by state-of-the-art numerical simulations}

\subsection{The dynamical atmosphere as seen by CO5BOLD}\label{sec_dynamics}

In this section, we concentrate on the results obtained with CO5BOLD to report the outcome of the convective cycles extracted from high spectral resolution spectroscopy (Section~\ref{sec_cycles}). We present in detail the dynamical processes emerging from the simulations of AGB and RSG stellar atmospheres (Table~\ref{simus}). 


\subsubsection{Starting a simulation}

As described in \cite{2011A&A...535A..22C},  the principal requirements that determine the type of the simulated star are (i) the input luminosity into the core, (ii) the stellar mass (i.e., the gravitational potential), (iii) the chemical abundances that affect the atomic physics through the opacity and equation of state tables. \\
Moreover, several additional parameters influence the outcome of a simulation to some extent: the formulation of the boundary conditions, the numerical resolution, the hydrodynamics scheme, the tensor viscosity, the choice of ray directions for the radiative transport, and the time-step control \citep{2012JCoPh.231..919F,2013MSAIS..24...26F}. One particular point concerns the closed-bottom boundary: as the radiative flux in the convection zone of cool stars with efficient convection is negligible, the energy in the bottom layers is not injected by radiation but is added as an energy source term within the central “internal” boundary. In these layers, the presence of strong velocity fields (waves or downdrafts) needs an adapted treatment of boundary conditions to avoid the generation of spurious acoustic waves when downdrafts bounce against a hard lower boundary. In addition to this, there are also options to set the boundary conditions handling the energy-source terms with the possibility to include a drag force in the core \citep{2017MmSAI..88...12F}. The (magneto-) hydrodynamics and radiation-transport solvers integrate right through this zone.

\begin{figure}
  \includegraphics[width=1.\hsize]{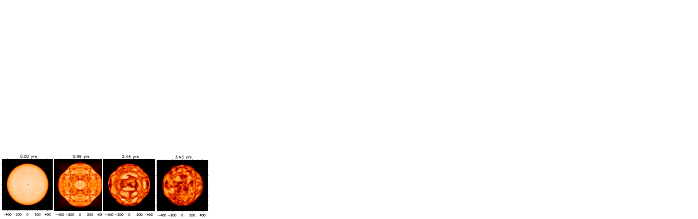}
\caption{Example of a starting RHD simulation for a RSG star \citep{2009A&A...506.1351C}. The panels show the evolution of the bolometric intensity from a sphere in hydrostatic equilibrium (leftmost panel) to the relaxed simulation (rightmost panel). The artifacts caused by the mismatch between the spherical object and the Cartesian grid become less evident with time.}
\label{fig:initialmodel}       
\end{figure}

The initial model is a sphere in hydrostatic equilibrium with a weak velocity field inherited from a previous model with different stellar parameters. The initial temperature stratification follows a gray atmosphere with the optically thick layers chosen to be adiabatic. Fig.~\ref{fig:initialmodel} shows the time evolution. The first snapshot in the leftmost panel does not display any convection-related structures but with some regular patterns due to the coarse numerical grid and poor temperature sampling at the sharp edge of the atmosphere. Later, a central spot appears and then vanishes completely when convection becomes strong. After several stellar years, the intensity contrast has grown and the surface pattern becomes completely irregular.  All memory from the initial symmetry is lost as well as the influence of the Cartesian grid. Once the simulation has reached a relaxed state, stellar parameter fluctuations stabilize and average quantities can be computed. 

\begin{figure}
  \centering
    \begin{tabular}{cc} 
    \includegraphics[width=0.9\textwidth]{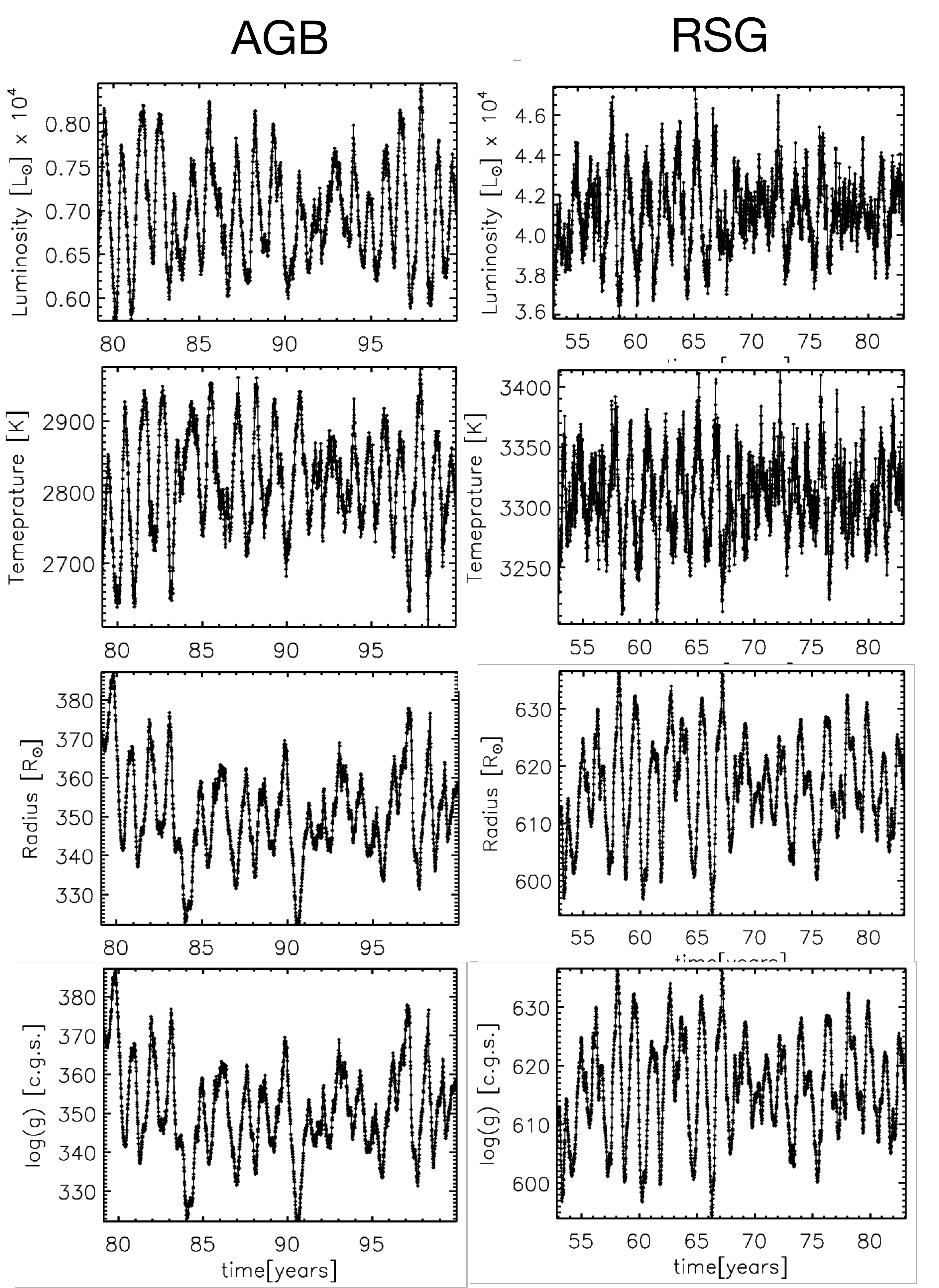}\\
\end{tabular}
\caption{Spherically averaged, global properties as a function of time for an AGB (Left-hand column) and an RSG simulation (Right-hand column). See Table~\ref{simus} for their temporal averages.}
\label{fig:stellarparam}       
\end{figure}

Table~\ref{simus} displays the temporal and spherically averaged stellar parameters and the numerical box details of two RHD simulations representing AGB and RSG stars, while Fig.~\ref{fig:stellarparam} shows the spherical averaged quantities still vary as a function of time as explained in \cite{2009A&A...506.1351C} and \cite{2011A&A...535A..22C}.

\begin{table*}[h!]
\begin{center}
 \caption{Parameters of the RHD simulations. The first two columns show the simulation name and stellar type. The following five columns display the stellar parameters such as the mass ($M_\star$) used for the external potential, the average emitted luminosity ($L_\star$), the Rosseland radius ($R_\star$), the effective temperature 
($T_{\mathrm{eff}}$) and surface gravity ($\log g$) at the Rosseland radius. Then the pulsation period $P_\mathrm{puls}$ and the full-width half-maximum (FWHM) of the distribution of pulsation frequencies $\sigma_{\mathrm{puls}}$. All these quantities are averaged over different epochs (9th column, $t_{\mathrm{avg}}$). Errors are one-standard-deviation fluctuations with respect to the time average \citep[see][]{2011A&A...535A..22C,2009A&A...506.1351C}. Solar metallicity is assumed. The simulation of the AGB comes from the grid of \cite{2017A&A...600A.137F} and it is used also in \cite{2021arXiv211210695C}. The simulation of the RSG has been computed including a rotational period of 30 years adding a centrifugal potential to the gravitational potential and the Coriolis force which is applied at each hydrodynamics step \citep{2017A&A...600A.137F}.}
 \label{simus}
 \resizebox{\textwidth}{!}{
 \begin{tabular}{|c|c|c|c|c|c|c|c|c|c|c|}
\hline
Simulation  &  $M_{\star}$ & $L_{\star}$ & $R_{\star}$ & $T_{\mathrm{eff}}$ & $\log g$ & $P_{\mathrm{puls}}$ & $\sigma_{\mathrm{puls}}$ & $t_{\mathrm{avg}}$ & Grid & x$_{\mathrm{box}}$\\
 & $[M_{\odot}]$  & $[L_{\odot}]$  & $[R_{\odot}]$ & [K] & [cgs] & [yr] &  [yr]  &  [yr]  & [grid points] & $[R_{\odot}]$ \\
\hline
st29gm06n001 & 1 & 6956.3$\pm$547.4 & 350.4$\pm$11.3 & 2814.7$\pm$68.9 & $-$0.65$\pm$0.03 &  1.150 & 0.314 & 25.35 & 281$^3$ & 1381 \\
 AGB                &   &                                 &                            &                              &                               &            &            &          &                &          \\
 \hline
st35gm04n43   & 5 & 41033.4$\pm$1850.1 & 616.2$\pm$8.4 &  3308.2$\pm$33.2 & $-$0.45$\pm$0.01 &  1.456 & 0.209 & 30.21 & 315$^3$ & 1626 \\
RSG                 &   &                                 &                            &                              &                               &            &            &          &                &          \\
\hline
 \end{tabular}}
 \end{center}
\end{table*}



\subsubsection{Global quantities}

\begin{figure}
  \centering
    \begin{tabular}{cc}
  \includegraphics[width=0.46\hsize]{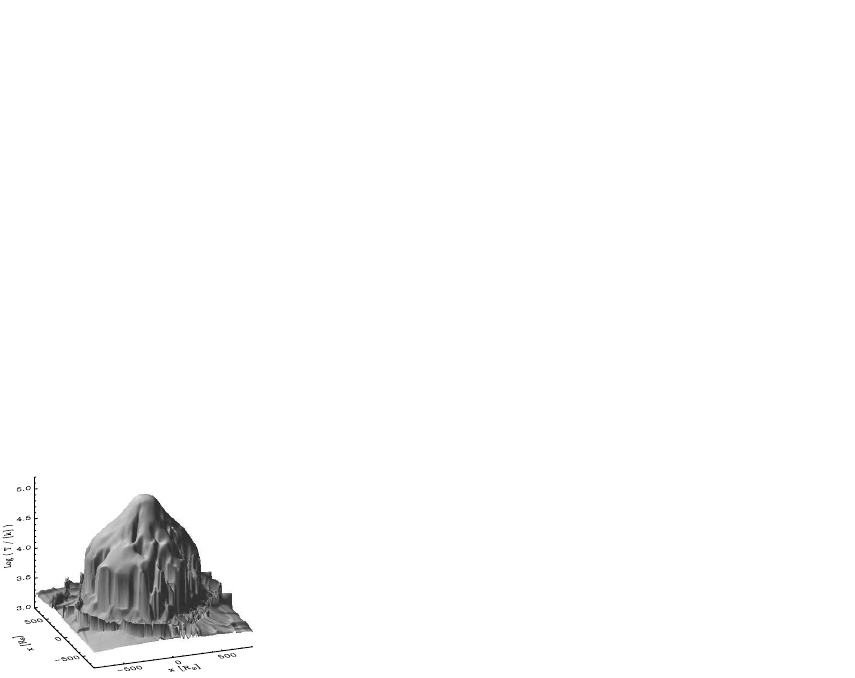}
    \includegraphics[width=0.46\hsize]{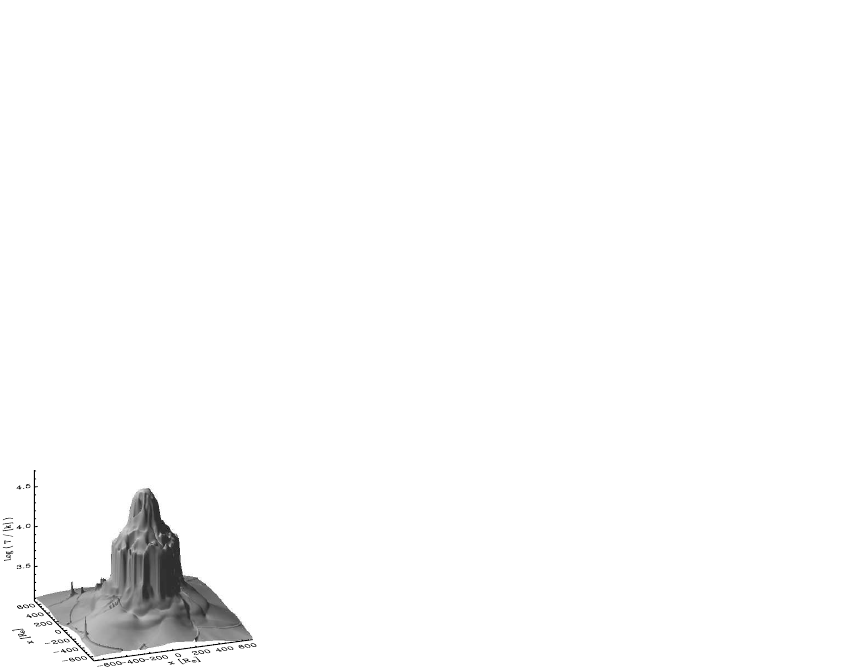}\\
     \includegraphics[width=0.46\hsize]{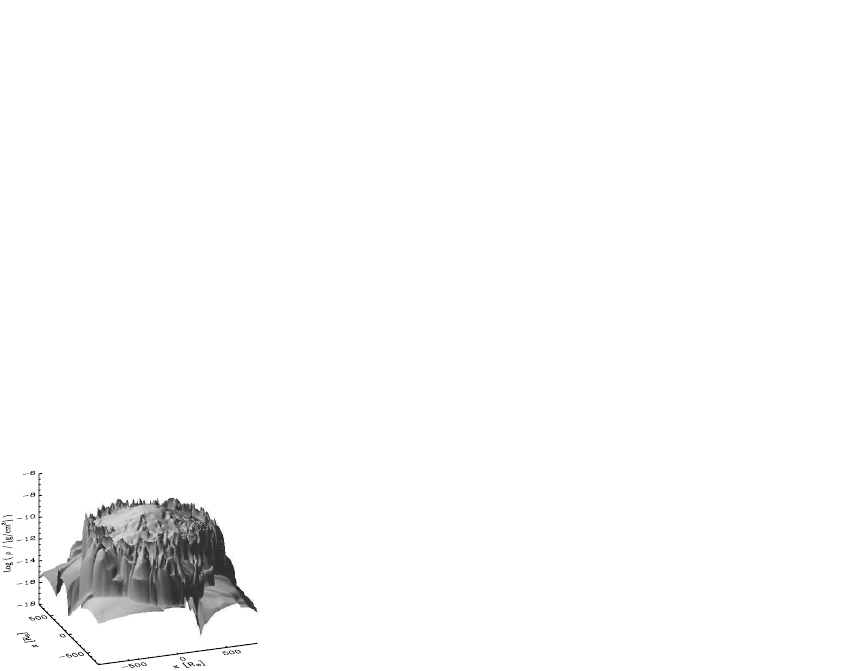}
    \includegraphics[width=0.46\hsize]{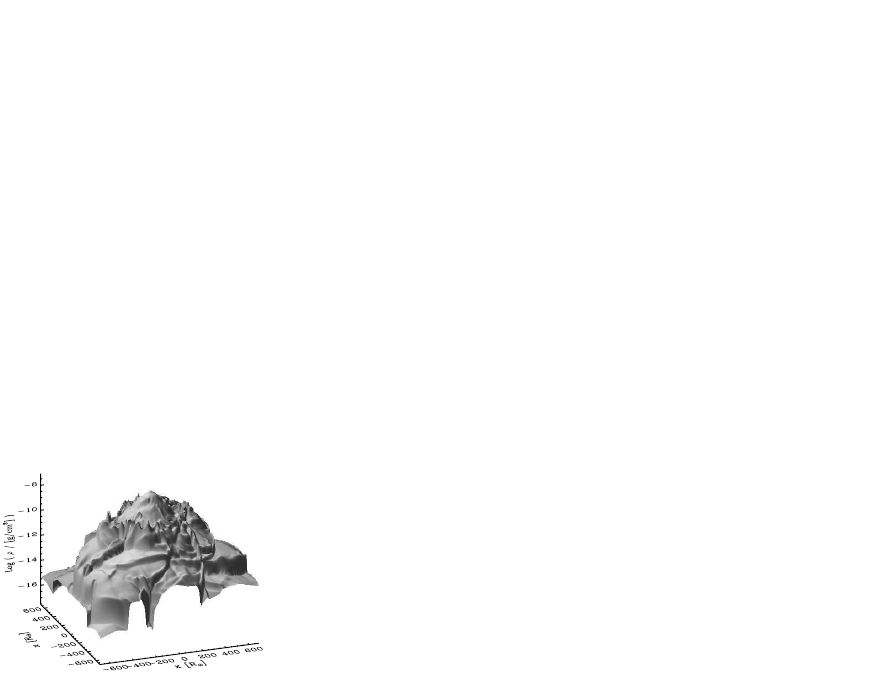}\\
      \includegraphics[width=0.46\hsize]{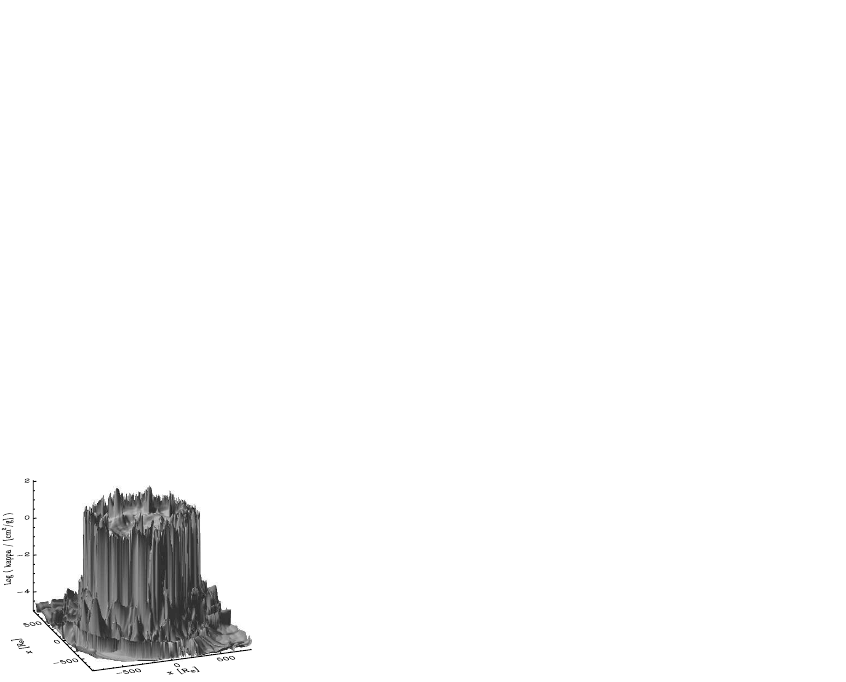}
     \includegraphics[width=0.46\hsize]{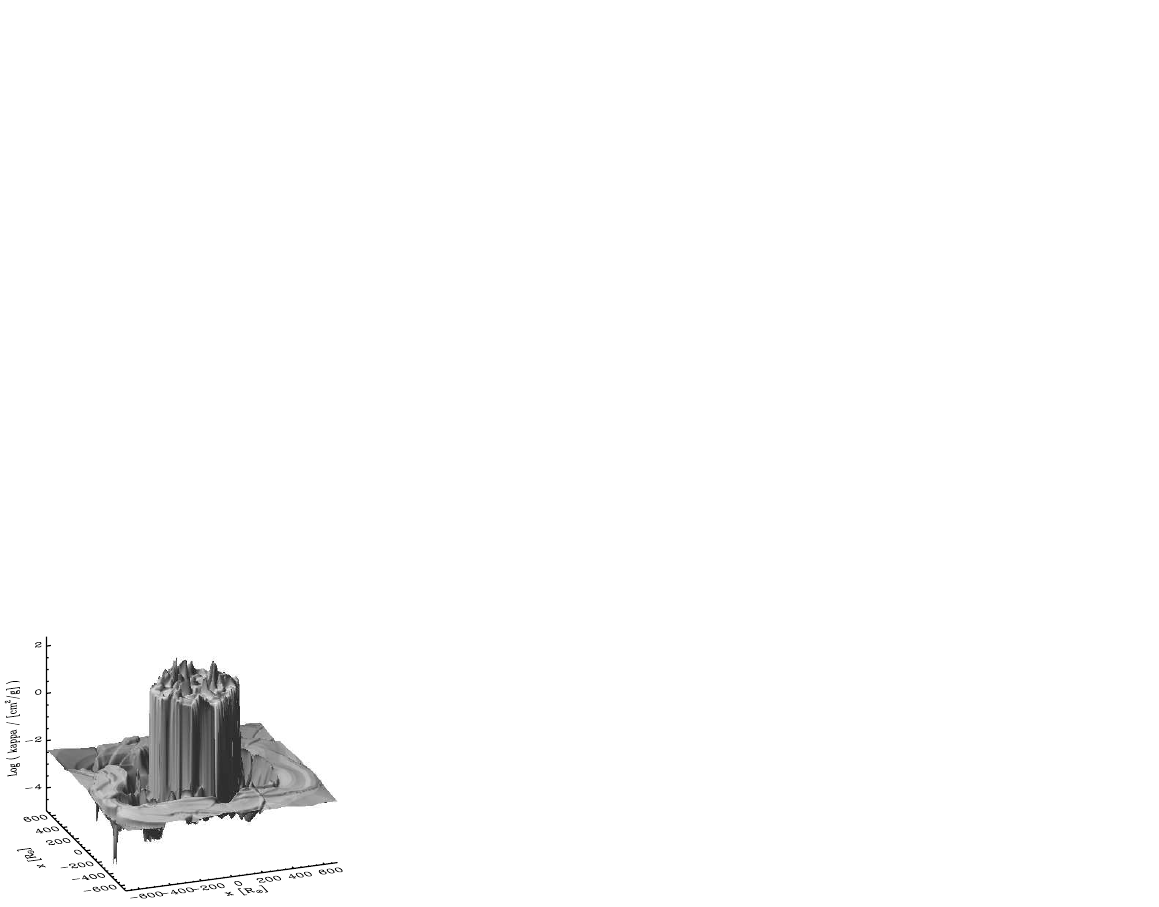}
          \end{tabular}
\caption{Logarithm of temperature (\emph{left column}), density (\emph{central column}), and opacity (\emph{right column}) of central slice of a snapshot for the RSG (left column) and AGB (right column) simulations of Table~\ref{simus}.}
\label{fig:globalquantities}       
\end{figure}

RHD simulations of evolved stars show a very heterogeneous surface caused by the dynamical granulation that evolves on timescales of weeks to years \citep{2011A&A...528A.120C,2017A&A...600A.137F,2023A&A...669A..49A}. 
 The radiation is of primary importance for many aspects of convection and envelope structure. Radiation cools down the surface to provide a somewhat unsharp outer boundary for the convective heat transport and it also contributes significantly to the energy transport in the interior, where convection never carries close to 100\% of the luminosity \citep{2011A&A...535A..22C}. Moreover, it must be noted that the entropy jump, happening in the optically thick region from the atmosphere to the layers below, is fairly large. \\
Below the visible layers (i.e., optical depth $\tau_{\mathrm{Rosselend}}>1$), the opacity has its peak at $T\sim13,000$ K (see the temperature and opacity surface rendering in Fig.~\ref{fig:globalquantities}) causing a very steep temperature jump which is prominent on top of upflow regions. At the same time a density inversion appears (see density in Fig.~\ref{fig:globalquantities}), which is a sufficient condition for convective instability. The entropy drop occurs in a very thin layer, while the smearing from the averaging procedure over nonstationary up- and downflows leads to the large apparent extent of the mean superadiabatic layer \citep{2011A&A...535A..22C}.

\begin{figure}
  \includegraphics[width=1.0\hsize]{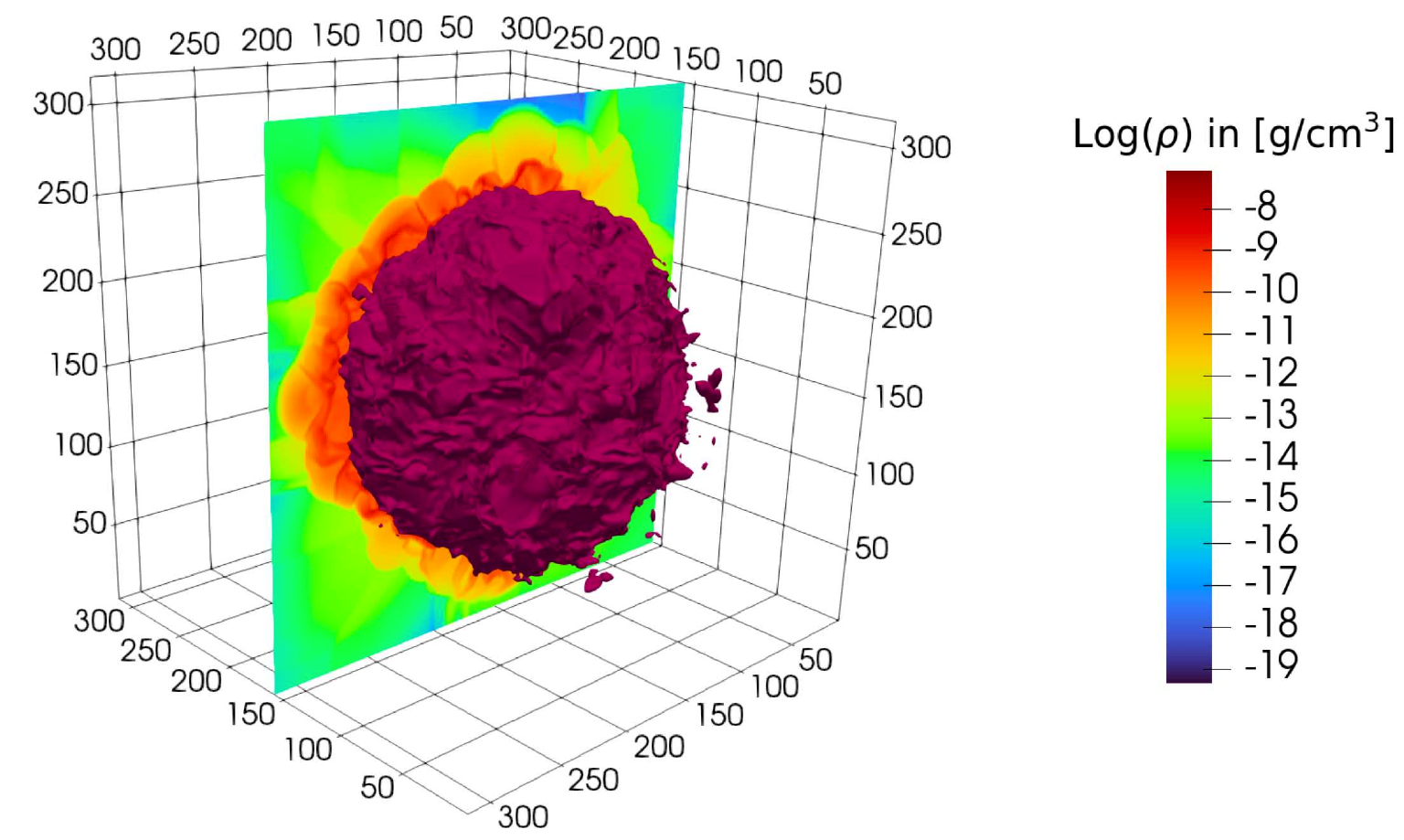}
\caption{Logarithm of the density (increasing from blue to red) in a slice through the center of the RSG simulation of Table~\ref{simus} overplotted to the isosurface of the Rosseland photosphere ($\tau_{\rm Ross}=1$), where the continuum flux is formed, is shown in amaranth color.}
\label{fig:density}       
\end{figure}

\begin{figure}
  \centering
    \begin{tabular}{cc}
  \includegraphics[width=0.45\hsize]{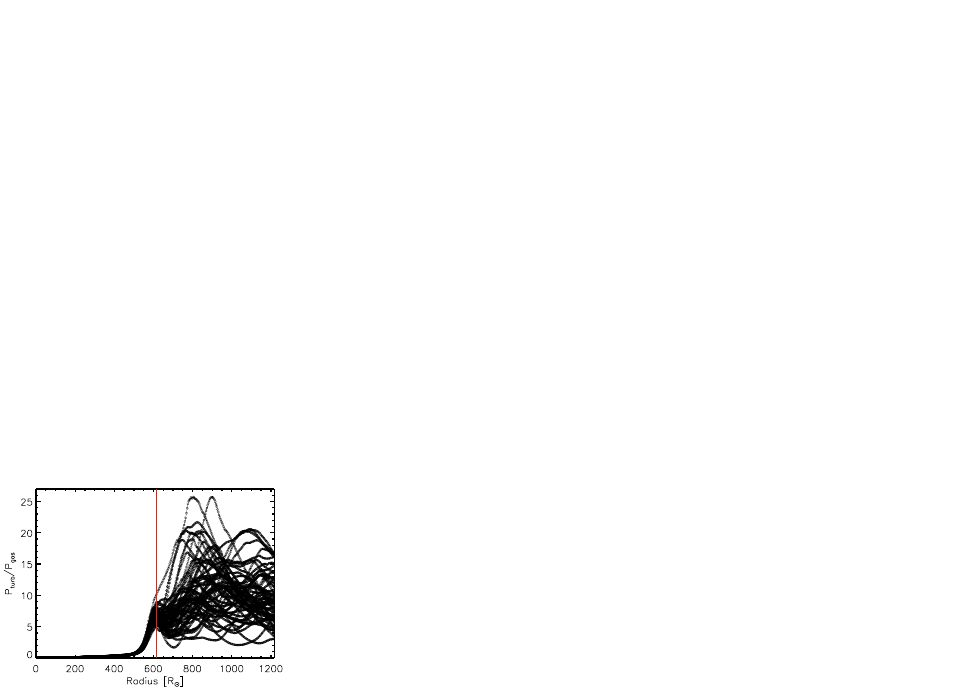}
   \includegraphics[width=0.45\hsize]{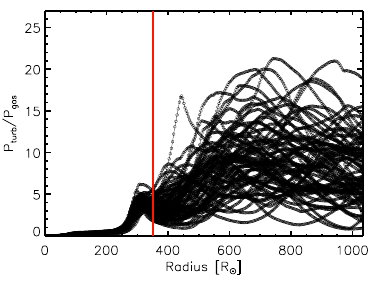}\\
     \includegraphics[width=0.45\hsize]{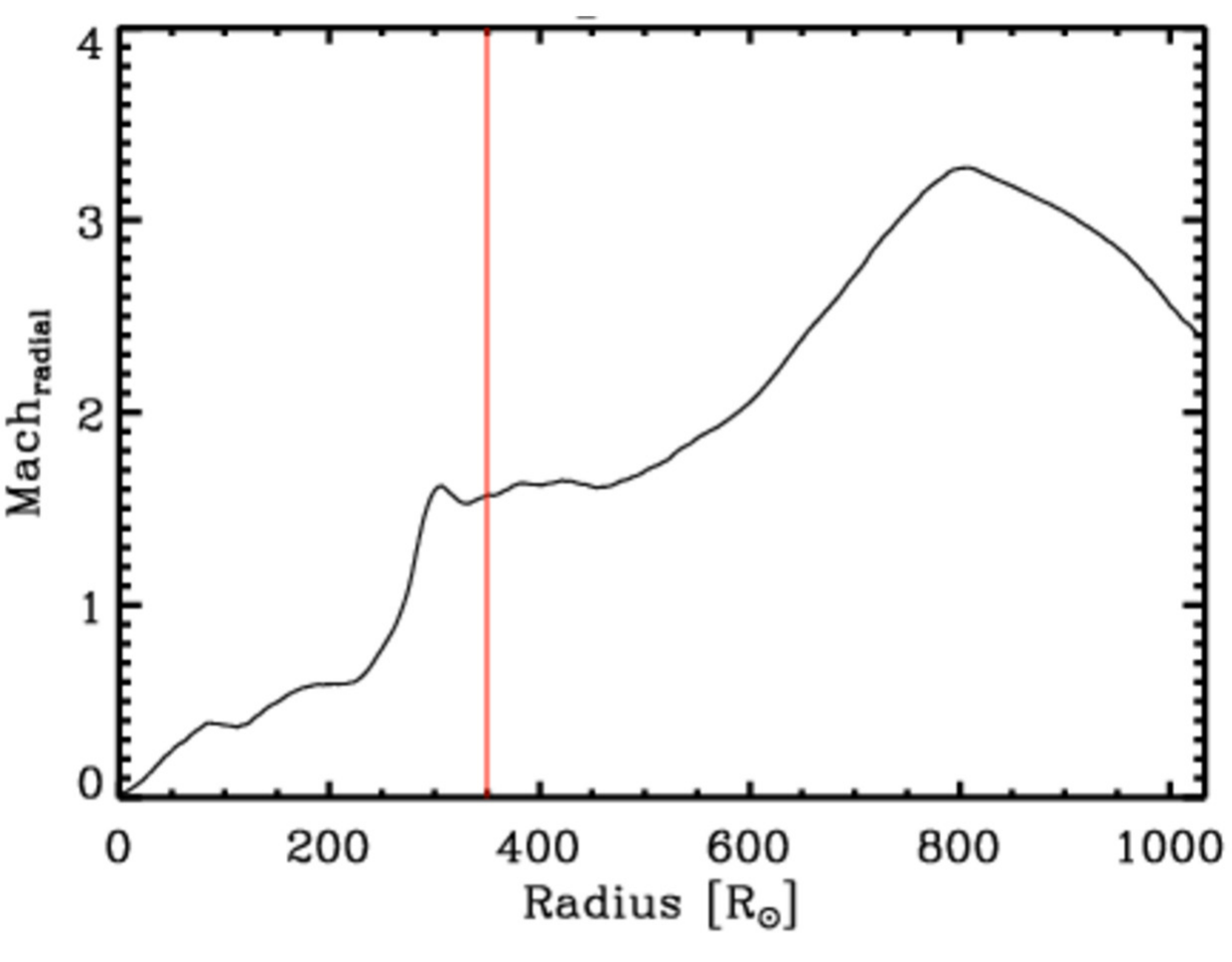}
   \includegraphics[width=0.45\hsize]{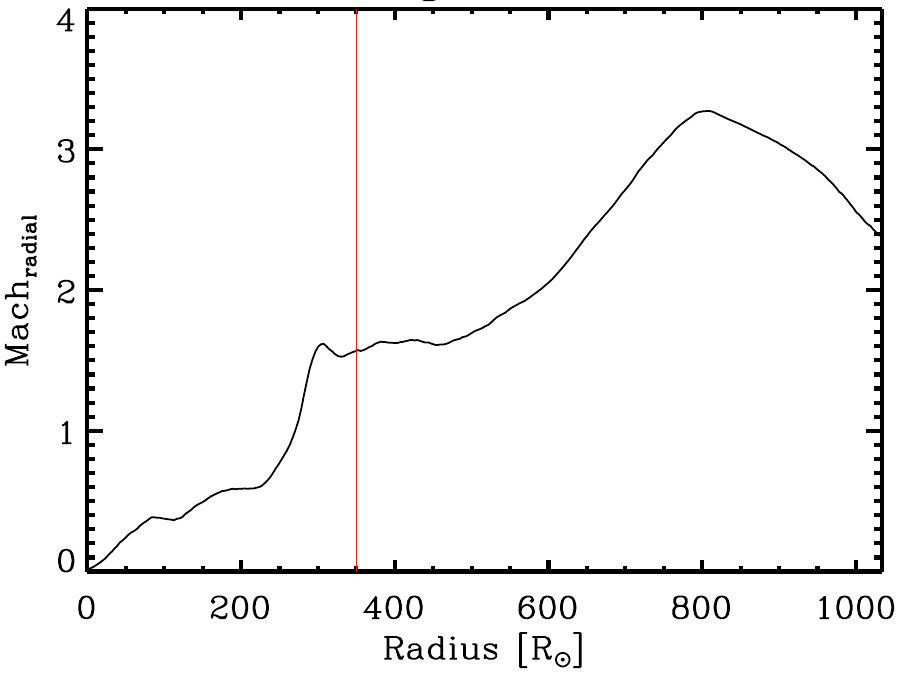}\\
          \end{tabular}
\caption{\emph{Top row:} ratio between turbulent pressure and gas pressure for different snapshots of the simulations. \emph{Bottom row:} spherical shells averages over time of the radial Mach numbers. 
The left panels show the RSG and the right one the AGB simulation. The red vertical lines in all the panels is the approximative position of the radius from Table~\ref{simus}.}
\label{fig:turbpress}       
\end{figure}

The emerging visible luminosity is related to layers where waves and shocks dominate together with the variation in opacity through the atmosphere. The global convective flow in the simulations is characterised by rising material that originates in the NB{convective zone  below the Rosseland radius} and develops as an atmospheric shock when it reaches higher values in radius \citep{2017A&A...600A.137F,2018A&A...619A..47L}. In deeper layers, non-stationary convection (e.g., merging down-drafts or other localized events) produces sound wave in the stellar interior and only a few large-scale cells dominate the convective flow. The sound waves travel through the stellar interior ($\tau_{\mathrm{Rosselend}}>1$) to the outer optically thin layers  where they hit the surface, slow down and get compressed due to the drop in temperature and sound speed. The amplitude of the sound wave rises due to the decrease in density (Fig.~\ref{fig:density}) and turns into a shock \citep{2019A&A...632A..28K}. Here, the turbulent pressure can be as high as more than 20 times the gas pressure (Fig.~\ref{fig:turbpress}, top row). Just above the stellar radius, the turbulent pressure contributes significantly to the average pressure stratification and to the radial (supersonic) velocities. 

\begin{figure}
  \includegraphics[width=1.\hsize]{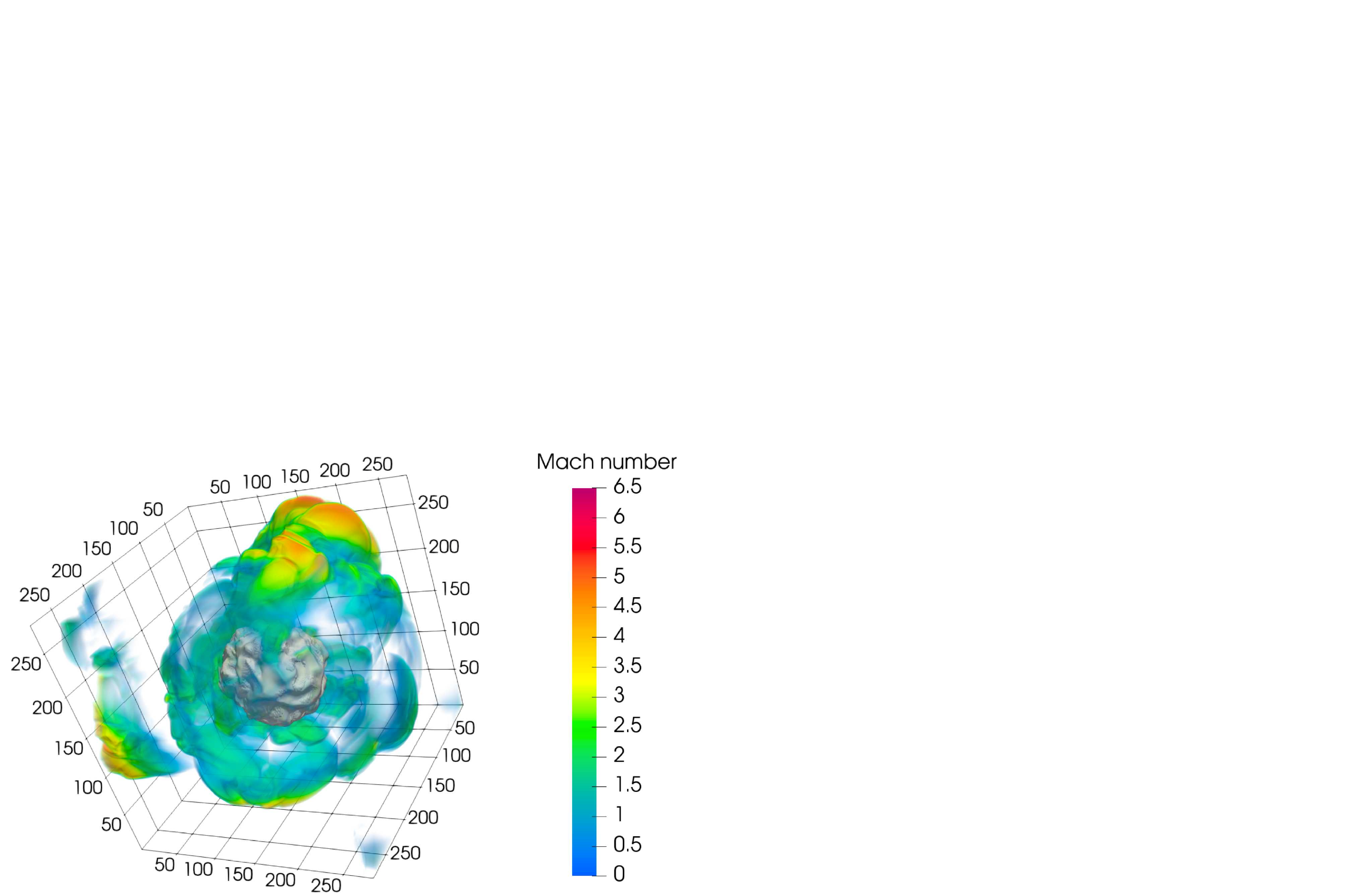}
\caption{Radial Mach numbers computed for the  RHD simulation of the AGB in Table~\ref{simus} (blue to red colors) rendered transparently over the isosurface of the temperature equal to 2800\,K (shaded, gray surface). This temperature corresponds approximatively to the effective temperature of the simulation (Table~\ref{simus}).}
\label{fig:machradial}       
\end{figure}

The newly formed shocks propagate from the stellar surface and out of the simulation domain with significant mach numbers of 3--4 on average (Fig.~\ref{fig:turbpress}, bottom row). These shock waves have lasting effects on the outer layers making them
     extremely heterogeneous and having Mach numbers peaking at 8--10. (Fig.~\ref{fig:machradial}). In these layers both the density and the temperature show irregular structures with convection cells in the interior and a network of shocks above the photosphere (Fig.~\ref{fig:density} and Fig.~\ref{fig:globalquantities}). Local fluctuations in high Mach numbers as well as small-scale heights due to shocks pose high demands on the stability for the hydrodynamic solver. A side effect of the steep and significant temperature jump is the increase in pressure scale height from small atmospheric values to values that are a considerable fraction of the radius in layers just below the atmosphere. The simulation in Fig.~\ref{fig:machradial} shows that the outer boundaries are either hit at some angle by an outgoing shock wave or let the material fall back (mostly with supersonic velocities). The shocks pass through the boundaries with a simple and stable prescription in the code based on filling typically two layers of ghost cells where the velocity components and the internal energy per unit mass are kept constant \citep{2012JCoPh.231..919F}.

\subsection{The convective envelope as seen by Athena++ }\label{sec_dynamics}

In this section, we focus on the Athena++ simulations which have been used to calibrate mixing length parameters and predict signatures of the 3D convective outer layers in the supernova shock breakout (Section~\ref{sec_AthenaResults}). We describe the stellar structure as seen in the simulations of two RSG envelopes (Table~\ref{tab:3Dmodels}) in higher- and lower- luminosity regimes. 

\subsubsection{Conducting a simulation in Athena++}
\begin{table*}\label{tab:3Dmodels}
\begin{center}
\resizebox{\textwidth}{!}{
\begin{tabular}{| c | c | c | c | c | c | c | c | }
\hline Name & $R_\mathrm{IB}/R_\odot$ & $R_\mathrm{out}/R_\odot$ & heat source & resolution ($r\times\theta\times\phi$) & duration & $m_\mathrm{IB}/M_\odot$ & $M_\mathrm{final}/M_\odot$ \\ \hline
RSG1L4.5 & 400 & 22400 & ``hot plate" & $384\times128\times256$ & 5865 d & 12.8 & $16.4$ \\ \hline

RSG2L4.9 & 300 & 6700 & fixed $L$ & $256\times128\times256$ & 5766 d & $10.79$ & $12.9$ \\ \hline
\end{tabular}}
\end{center}
\caption{Simulation properties of the Athena++ models from \citet{Goldberg2022a}, including inner boundary radius ($R_\mathrm{IB}$), outer boundary ($R_\mathrm{IB}$), heat source, resolution, run duration, core mass $m_\mathrm{IB}$, and total mass at the simulation end ($M_\mathrm{final}$). All models have $\theta=\pi/4-3\pi/4$ and $\phi=0-\pi$, with $\delta r/r\approx 0.01$. The naming scheme indicates the log of the time-averaged luminosity $\log(L/L_\odot)$.}
\end{table*}

The Athena++ Red Supergiant setup described in \citet{Goldberg2022a} is based on the semi-global (large solid-angle) ``Star3D" setup presented in \citet{2018Natur.561..498J}, with the difference that the gravitational potential accurately includes the (non-negligible) time-dependent, spherically symmetric, envelope mass, $g(r) = -\nabla\Phi(r) = G(m_{\rm IB} + m_{\rm env}(r))/r^2$ where $m_{\rm env}(r)$ is the mass interior to the shell at radius $r$. These simulations cover 70.6\% of the face-on hemisphere (i.e. solid angle $\Omega=1.41\pi$) using spherical polar coordinates with 256 uniform bins in azimuth $\phi$ from $0-\pi$  and 128 uniform bins in polar angle 
$\theta$ from $\pi/4-3\pi/4$, with periodic boundary 
conditions in $\theta$ and $\phi$ enforced by ghost zones outside the simulation domain \citep[see][]{2020ApJS..249....4S} to conserve mass and energy flux and to avoid the ``splashback" phenomenon associated with convective motions through reflective boundaries, and logarithmic bins in radius with $\delta r/r\approx0.01$. 

\begin{figure*}
\centering
\includegraphics[width=\textwidth]{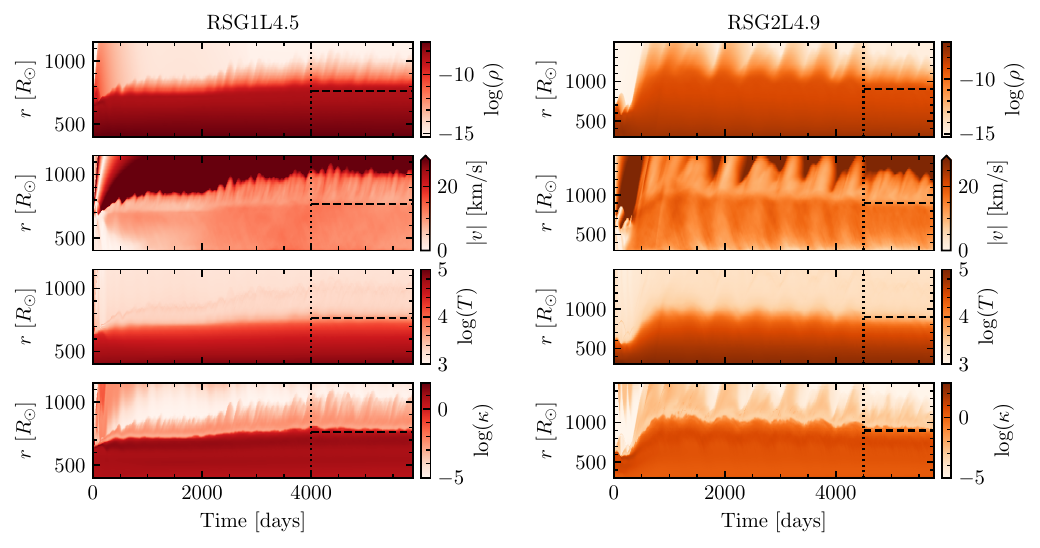}
\caption{History of the averaged radial profiles for the RSG1L4.5 (red, left) and RSG2L4.9 (orange, right) Athena++ models. Top to bottom show log(density), magnitude of the 3D velocity, log(temperature), and log(opacity). All logarithms are base 10 and units inside the logarithms are cgs, and velocites are reported in km/s. Vertical dotted lines indicate when the envelope appears to have reached a convective steady state. Horizontal dashed lines approximate the region where some fraction of the stellar area has $\tau>\tau_{\rm crit}$. Figure from \citet{Goldberg2022a}.}
\label{fig:ssprofiles}
\end{figure*}

The inner boundary (IB) has fixed radius at $400R_\odot$ and $300R_\odot$ for RSG1L4.5 and RSG2L4.9 (named by the log of the time-average luminosity leaving the simulation domain) respectively with a zero-velocity, fixed temperature, fixed density inner boundary condition, and analysis is restricted to discussing the convective region which exhibits steady-state behavior, from  $\approx$ 450$R_\odot$ outwards. 
The implicit radiation scheme implemented in Athena++ \citep{Jiang2014,Jiang2021} is used, which directly solves the RHD equations for specific intensity over discrete ordinates, including radiation in both the energy and momentum equations as a source term. The gas equation of state is taken to be a gamma-law where $\gamma=5/3$, as the radiation pressure is included explicitly. For the two models discussed, initial conditions are constructed in radiative equilibrium from these values at the inner boundaries which were motivated by 1D models constructed using the stellar evolution code MESA \citep{Paxton2011,Paxton2013,Paxton2015,Paxton2018,Paxton2019, Jermyn2023}. 
The luminosity is then either supplied by a ``hot plate" method (RSG1L4.5), where the temperature is increased at the inner boundary to drive convection, or a fixed high L coming in at the inner boundary (RSG2L4.9). 

The simulations are run until they achieve a convective steady state as discussed in more detail in \citealt{Goldberg2022a}. This time is shown by the vertical dotted line in Fig.~\ref{fig:ssprofiles}, which presents the time-evolution of the radially-averaged density, velocity, temperature, and opacity for the two Athena++ models. 
Also indicated by the horizontal dashed line is the location where radiation is expected to carry a significant fraction of the stellar flux, 
$\tau \approx \tau_{\rm crit}$ where $\tau_{\rm crit} \approx P_{\rm rad}/P_{\rm gas} c/|v_r| ~$ is a few hundred in the RSG regime (e.g. \citealt{Goldberg2022a,2022RNAAS...6...29J,Schultz2023}).

\subsubsection{Envelope Structure and Convective Properties in Athena++}

Like the CO5BOLD models discussed above, the Athena++ RHD simulations show very heterogeneous outer layers with convective overturn timescales on the order of a year. 
\begin{figure}
\centering
\includegraphics[width=1.\textwidth]{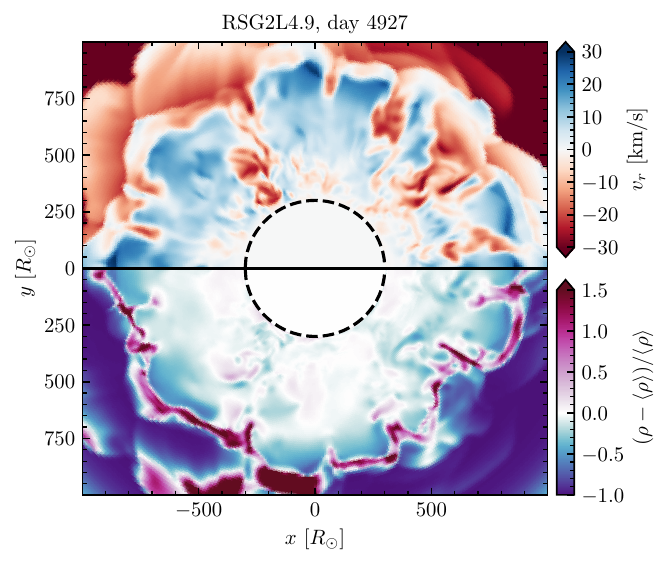}
\caption{ \label{fig:vdrho} Equatorial ($\theta=\pi/2, z=0$) slice of a characteristic snapshot of the RSG2L4.9 model, showing (upper half of the figure:) radial velocities and (lower half of the figure:) density fluctuations relative to the shellular volume-averaged density at each radius. The simulation domain is from $\phi=0$ to $\pi$; thus the image is reflected about $y=0$ as indicated by the axis labels. Figure from \citet{Goldberg2022a}.}
\end{figure}
\begin{figure}
\centering
\includegraphics[width=1.\textwidth]{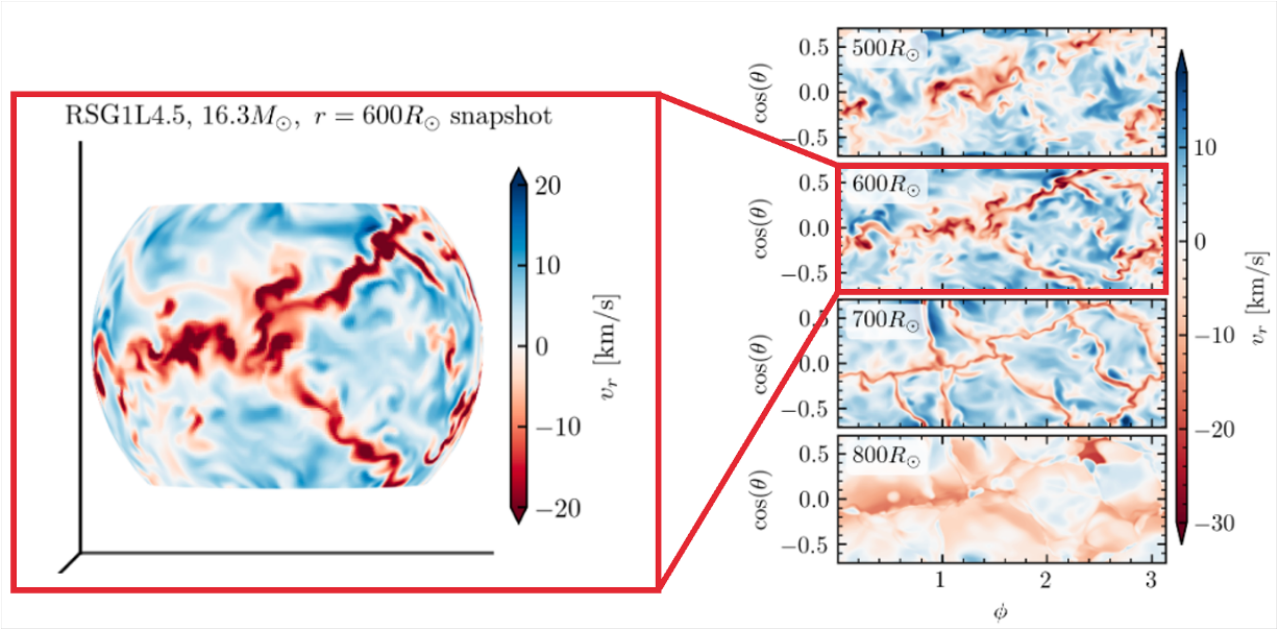}
\caption{Left: Surface rendering of the radial velocity fluctuations at $r=600R_\odot$ in RSG1L4.5 at day 4707. Right: Snapshot of radial velocity slices for the same model. Panels show radial slices at $r=500R_\odot$ (top) to $800R_\odot$ (bottom) in $100R_\odot$ intervals, and axes show the extent in azimuth $\phi$ and co-polar angle cos($\theta$).
The volume-weighted mean sound speed is 33\,km/s at 
$r=500R_\odot$, 26\,km/s at $600R_\odot$, 19\,km/s at $700R_\odot$, and 8\,km/s 
at $800R_\odot$. Figure adapted from \citet{Goldberg2022a}.}
\label{fig:vradial}
\end{figure}

The radial and convective velocities in the Athena++ models remain coherent on large scales throughout the convective envelope (see Fig.~\ref{fig:vradial}), in broad agreement with the flow morphology of the CO5BOLD simulations (e.g., recent work by \citealt{2023arXiv231116885M}). The density fluctuations are $\approx10\%$ in the middle of the convective envelope, reaching unity and greater at fixed radius in the outer envelope, as seen in Fig.~\ref{fig:vdrho} for RSG2L4.9. Although the net angular momentum in the envelope is nearly zero, at any given time the large-scale tangential velocity fluctuations (tens of km/s, coherent across tens to hundreds of solar radii) result in finite specific angular momentum ($j_\mathrm{rand}\sim10^{18}-2\times10^{19}$cm$^2$/s) at each radius, consistent with the recent 3D hydrodynamic simulations of interior RSG convection \citet{Antoni2022}.

\begin{figure}
\centering
\includegraphics[width=0.48\textwidth]{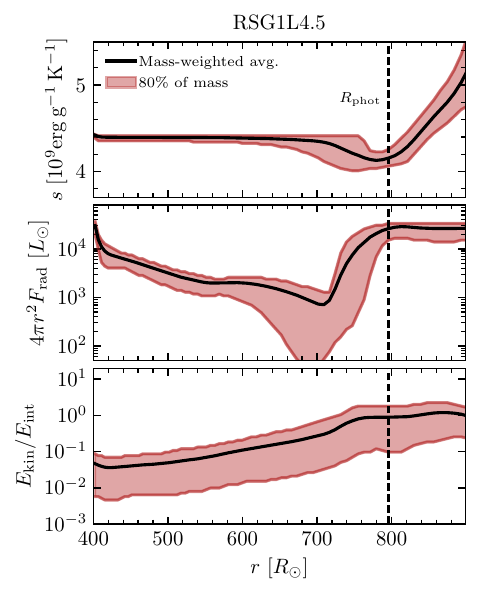}
\includegraphics[width=0.48\textwidth]{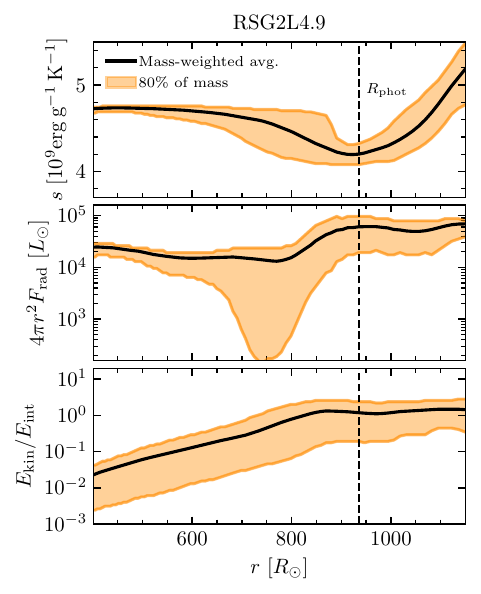}
\caption{ \label{fig:SLPratio} Specific entropy (top panels), radiative luminosity (middle panels), and ratio of the turbulent kinetic energy to the internal energy (bottom panels) in the RSG1L4.5 model at day 4707 (red, left) and RSG2L4.9 at day 4927 (orange, right). Mass-weighted averages are shown in black, with 80\% of the mass lying within the shaded regions. The 1D photosphere, where $\langle{L(r)}\rangle=4\pi r^2\sigma_\mathrm{SB}\langle{T_r(r)}\rangle^4$, is given by the vertical dashed line. Figure from \citet{Goldberg2022a}.} 
\end{figure}

Fig.~\ref{fig:SLPratio} shows profiles of the entropy, radiation flux, and kinetic energy to thermal energy ratio in the Athena++ RSG models of \citet{Goldberg2022a}. As radiation is included in both the energy and momentum equations in the Athena++ setup, the simulations achieve a flat entropy profile in the interior, a signature of efficient convection (as predicted by 1D models and MLT) which decreases and steepens as the convection becomes increasingly super-adiabatic. Radiation only carries $10-20\%$ of the flux in the interior, with convection carrying the rest. In the near-atmospheric layers, turbulent mach numbers in the vigorous, inefficient convection cross unity and the kinetic energy density matches or exceeds the thermal energy content, with turbulent pressure contributing significantly to hydrostatic balance. 

\begin{figure*}
\centering
\includegraphics[width=0.95\textwidth]{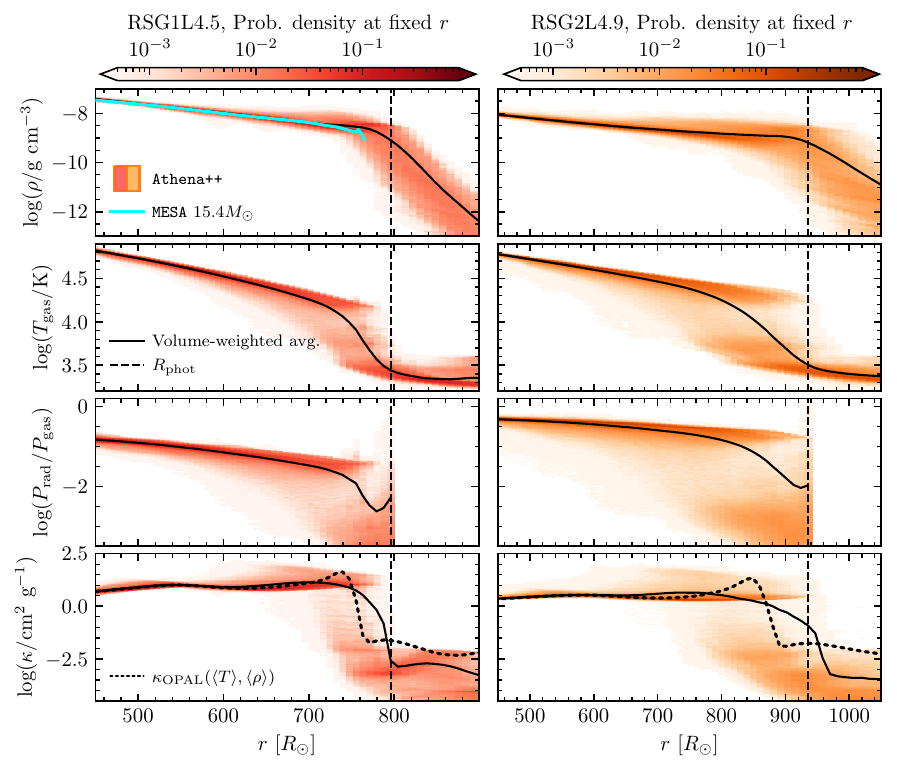}
\caption{Top to bottom: Density, temperature, 
$P_{\rm rad}/P_{\rm gas}$ ratio, and opacity for the RSG1L4.5 model at day 4707 
(red, left panels) and the RSG2L4.9 model at day 4927 (orange, right 
panels). Color saturation indicates the volume-weighted probability of 
finding a fluid element at a given ($\rho, T,$ pressure scale height $H/r, P_{\rm rad}/P_{\rm gas}, \kappa$)
at each radial coordinate. Solid black lines give
the volume-weighted averages of each (non-log) quantity ($\langle\rho\rangle, \langle T\rangle, \langle P_{\rm rad}/P_{\rm gas}/\rangle, \langle\kappa\rangle$).
The vertical black dashed line is the photosphere radius $R_{\rm phot}$.  The ratio $P_{\rm rad}/P_{\rm gas}$ is only shown for 
$r\leq R_{\rm phot}$. The $\kappa$ panels (bottom) show both the volume-averaged 
opacity $\log(\langle\kappa\rangle)$ reported by Athena (solid lines), as well as the OPAL opacity (dotted lines) recovered by inputting the volume-averaged $T$ and $\rho$ profiles. For reference, the cyan line in the upper left panel shows the density profile of the MESA model which motivates the inner boundary and simulation initial conditions. Figure from \citep{Goldberg2022a}.}
\label{fig:3Dpanels}
\end{figure*}

The fluctuations in the stellar properties, which increase radially outwards in the envelope, can be seen in Fig.~\ref{fig:3Dpanels}.
The ratio of radiation to gas pressure is also significant in the interior, with 
$\langle P_{\rm rad}/P_{\rm gas}\rangle\approx0.15 - 0.5$ at the $r=450R_\odot$ coordinate, decreasing to the level of a few percent approaching the photsphere. 
Rather than a smooth transition from the H opacity
peak to scattering and neutral opacities, the temperature and opacity show
bimodal behavior (not present in the density profiles) in the $\approx100-200R_\odot$ region beneath the photosphere for both models. Thus the sharp opacity drop seen in 1D models which use only the volume-averaged $\rho$ and $T$ profiles appears less steep in the 3D simulations, though it is equally sharp along any given line-of-sight, as seen also in the CO5BOLD simulations.

Notably, the opacity $\kappa_\mathrm{Edd}$ above which $L$ locally exceeds the Eddington Luminosity $L_{\rm Edd}$, $\kappa_\mathrm{Edd}=4\pi Gc M/L$, is $\approx1-10$ cm$^2$/g, which means that a significant amount of the material at those transitionary radii with $\kappa\sim$~tens of cm$^2$/g~$>\kappa_\mathrm{Edd}$. 
At optical depths with radiation-dominated energy transport where some material exists at $\tau_{\rm crit}> \tau >1$, radiation forces may significantly impact fluid motion with high $\kappa\gtrsim \kappa_{\rm Edd}$, and MLT-like convection is not expected. In fact in those layers of the Athena++ simulations, the typical correlations between the radial velocity and the density, opacity, and entropy flip; such that the cold (low-entropy), opaque, dense regions are correlated with outward velocities, rather than simply sinking like they would in MLT-like convection.  Similar inverted-correlation 
behavior is also seen in other simulations of luminous stars (e.g. in OB-star envelopes; \citet{2022ApJ...924L..11S}), but not in simulations of solar-like convection (e.g. \citealt{1998ApJ...499..914S}),
likely owing to RHD effects where $\tau_{\rm crit}\gg1$ and $L\gtrsim L_{\rm Edd}$. 

\section{Results and applications}

\subsection{CO5BOLD applications: Convection cycles in evolved stars}\label{sec_cycles}

\cite{2017A&A...600A.137F} and \cite{2023A&A...669A..49A} reported that   convection and pulsations are both emergent properties of the RHD simulations. Moreover, \cite{2023A&A...669A..49A} showed the non-linear nature of the interplay between them: in all of their 3D RHD simulations, the radial pulsations can be observed and the pulsation periods in the fundamental mode extracted, with values in agreement with the observed periods of long period variable (LPV) stars.
 The authors also highlighted the importance of the characteristics of deep convective regions and large convection cells, for understanding the interplay with pulsations.\\
In this Section, we concentrate on the convection cycles based on the observational approaches related to visual magnitude and radial velocities measurements. Measurements of radial velocities are fundamental to determining stellar space velocities to investigate the kinematic structure of stellar populations in the Galaxy or to monitor for radial velocity variations, either of which could point to the presence of an unseen companions. In addition to this, radial velocities can also be used to recover the projected velocity field at different optical depths in the stellar atmosphere using the tomographic method described in the Section~\ref{sectiontomography}. 

In this context, convection plays an important role in the formation of observed spectral lines and deeply influences the spectral line formation in late-type stars \citep{2000A&A...359..669A}. Each line, forming across a range of depths in the atmosphere, has a unique fingerprint in the spectrum that depends on line strength, depth, shift, width, and asymmetry across the granulation pattern depending on their height of formation and sensitivity to the atmospheric conditions of main sequence and up to red giant branch stars \citep{2013A&A...550A.103A,2018A&A...611A..11C}. The RHD simulations provide a self-consistent ab-initio description of the non-thermal velocity field generated by convection, shock waves, and overshoot that manifests itself in spectral line shifts and changes in the equivalent width \citep{2011A&A...535A..22C,2015ASPC..497...11C} as revealed in the observations (e.g. \citealt{2008AJ....135.1450G}).

Finally, the velocity field in the evolved stars, reconstructed using the tomographic method (Section~\ref{sectiontomography}), can be related to the photometric variability to reveal phase shifts resulting in a hysteresis loop (Section~\ref{section_hysteresis}. According to \cite{2008AJ....135.1450G}, the hysteresis loop illustrates the convective turn-over of material in the stellar atmosphere: first, the rising hot matter reaches upper atmospheric layers, then the temperature drops as the matter falls back into the star. \\
In the following part of this Section, we present a detailed discussion of the methodology and possible origin of the hysteresis loop in evolved cool stars, taking into account the mechanisms related to the non-stationary effects of convection.

\subsubsection{Pulsation properties}

\begin{figure}
  \centering
    \begin{tabular}{c}
  \includegraphics[width=0.9\textwidth]{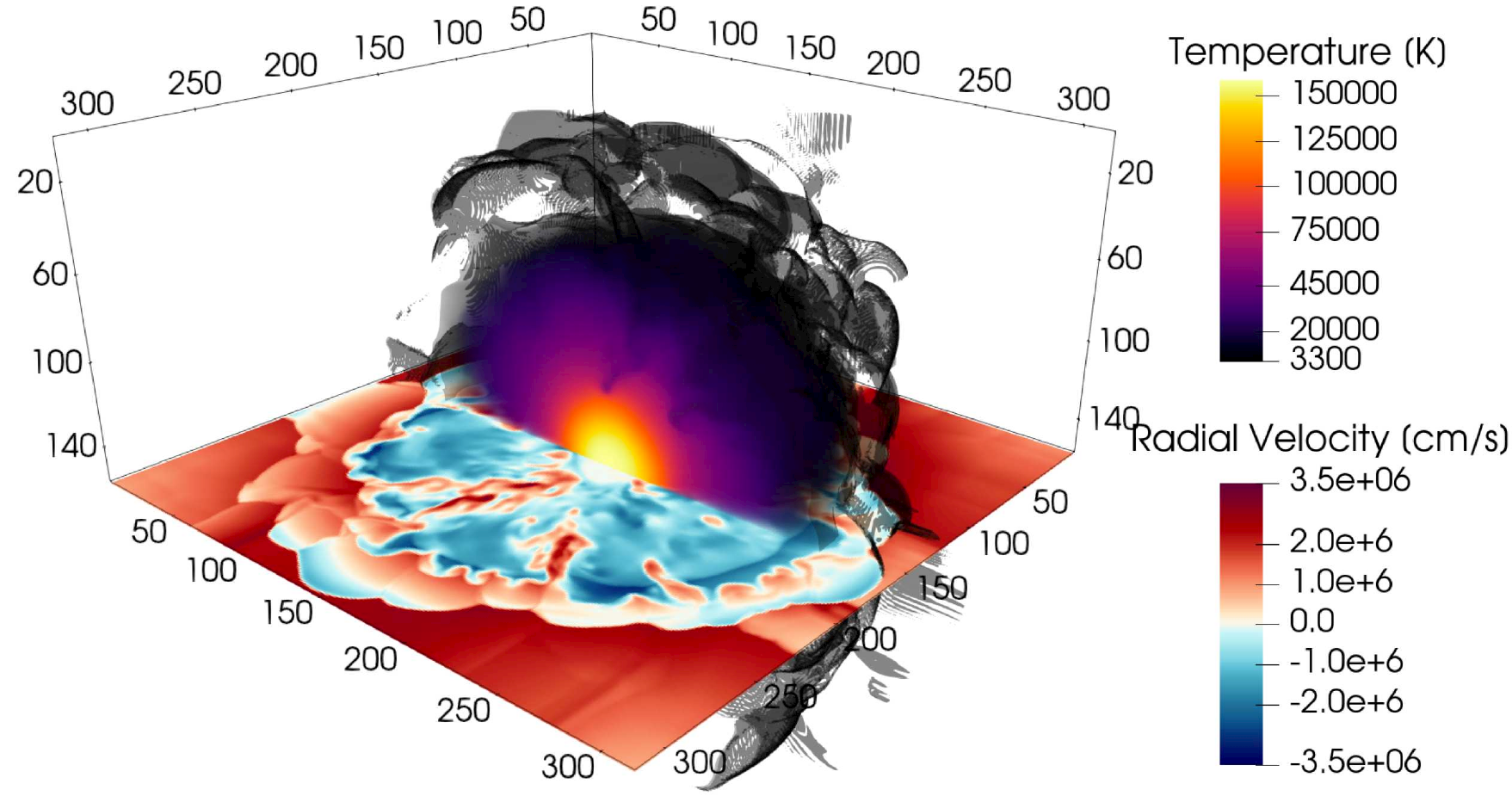}\\
    \includegraphics[width=0.85\textwidth]{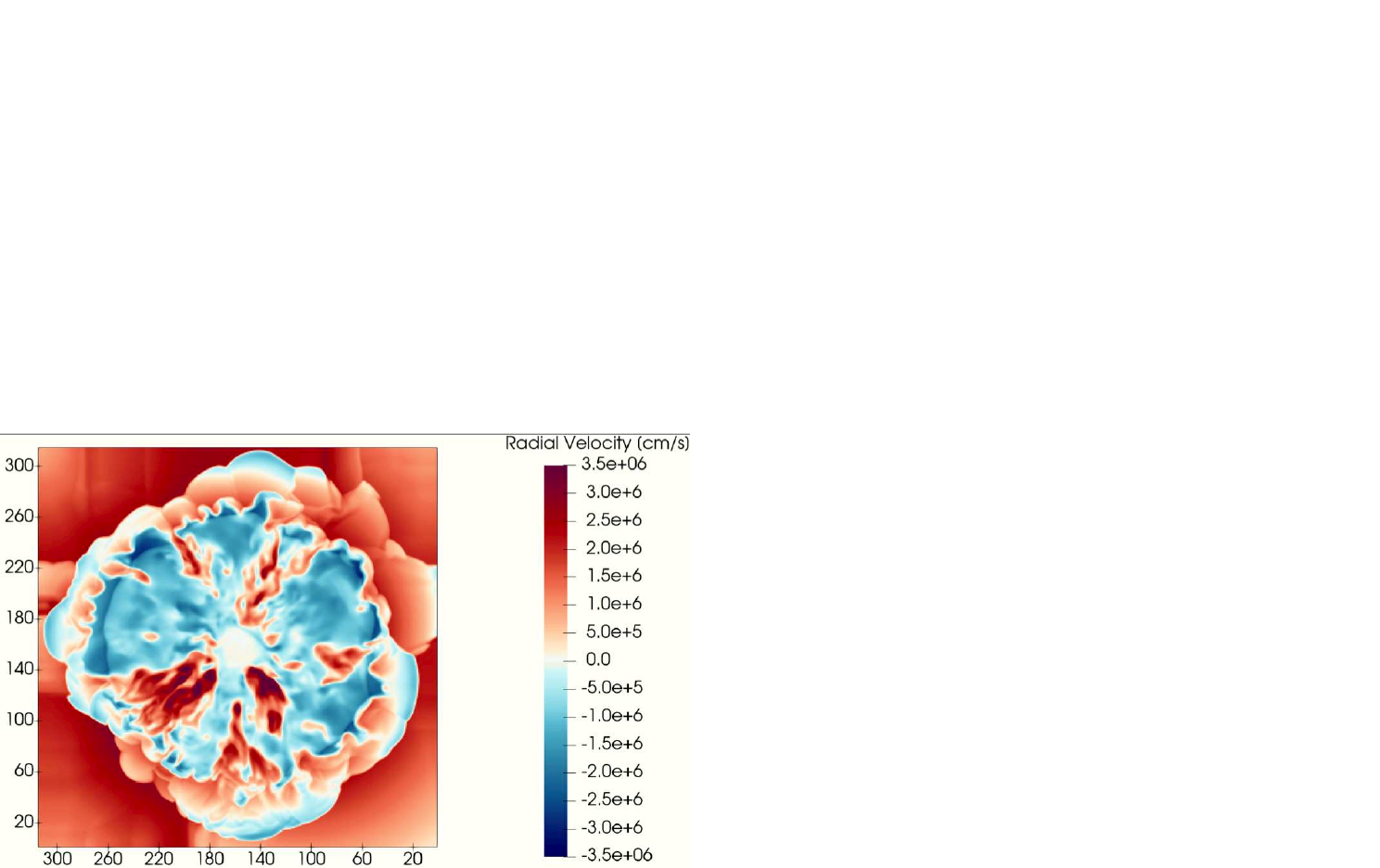}
          \end{tabular}
\caption{\emph{Top panel: }slice across the numerical box showing the radial velocity for the RSG simulation of Table~\ref{simus}. The blue indicates outward and red inward flow. Moreover, the temperature volume rendering is also displayed (yellow to black colors). The temperature values arbitrarily stop at the effective temperature of the simulations to show the approximate position of the $\tau_{\mathrm{Rosselend}}=1$. In the simulations the temperature can go down to $\sim$1000\, K. \emph{Bottom panel: } Enlargement of the radial velocity slice from above.}
\label{fig:radialvel1}       
\end{figure}

\begin{figure}
  \centering
    \begin{tabular}{cc}
\includegraphics[width=0.9\textwidth]{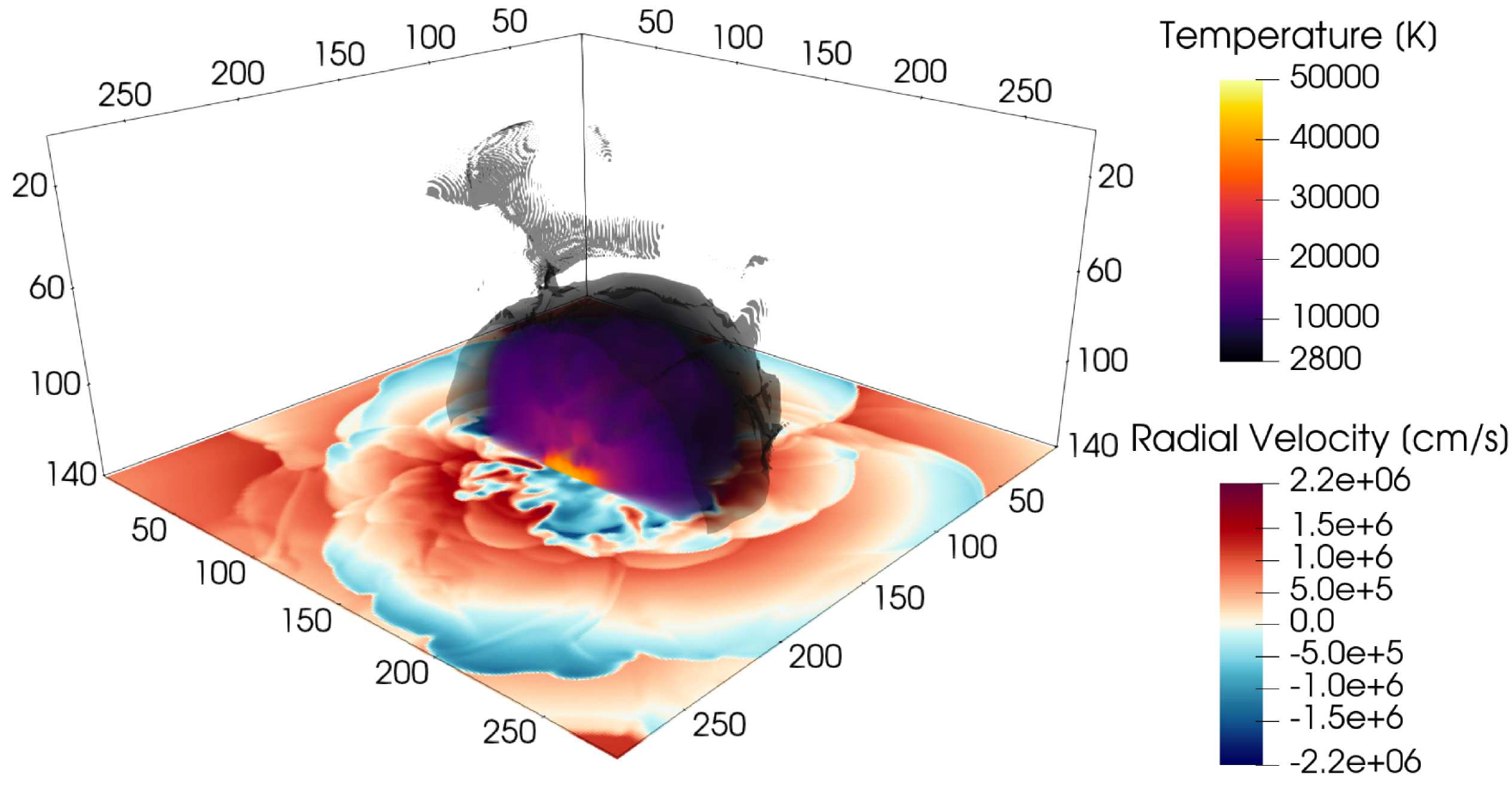}\\
    \includegraphics[width=0.85\textwidth]{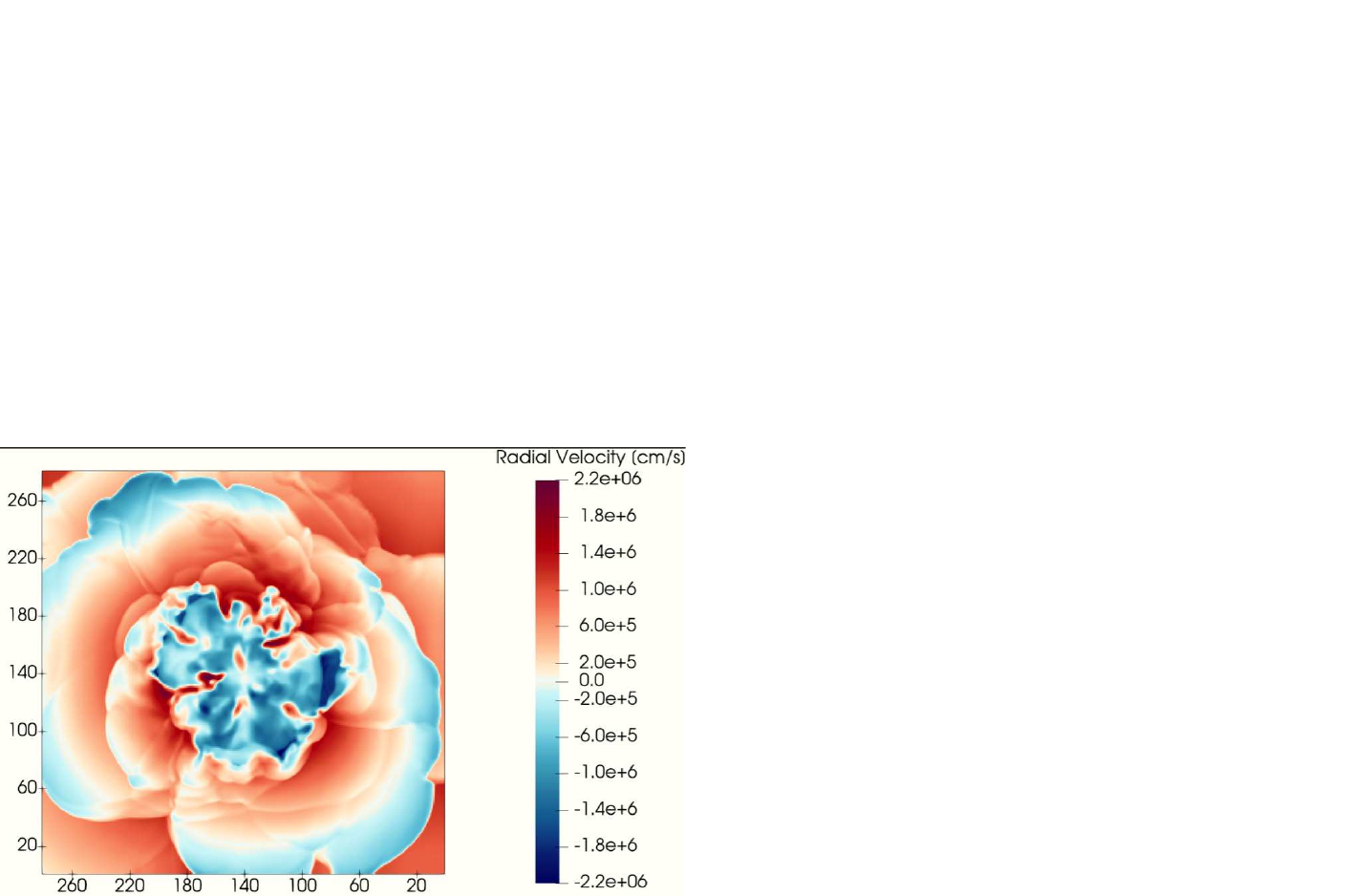}
          \end{tabular}
\caption{Same as in Fig.~\ref{fig:radialvel1} for the AGB simulation of Table~\ref{simus}.}
\label{fig:radialvel2}       
\end{figure}

The irregular structures of convection cells in the interior are visible in the temperature and density structures (Fig.~\ref{fig:globalquantities}) as well as in the radial velocity maps (Fig.~\ref{fig:radialvel1} and \ref{fig:radialvel2}) together with a network of shocks in the atmosphere of both the RSG or the AGB simulations. The high and heterogeneous velocities (up to $\sim$35 km/s in RSG simulation) and pressure fluctuations determine the acceleration of shock waves and, eventually, affect the mass loss. For the RSG, the large shocks just above $\tau_{\mathrm{Rosselend}}$ =1 can be $\sim200 - 250\, R_\odot$ wide  (each grid point is about $5\, R_\odot$, Table~\ref{simus}) for both the RSG and AGB with a local temperature of $\sim$2500\,K and 
$\log(\rho/[{\rm g\,cm}^{-3}])\sim -13$. This width is comparable to the actual radius of the star. \\
The rising material (blue color) observed in the figure originates in the deep convective zone and an atmospheric shock develops higher up ($\tau_{\mathrm{Rosselend}}$ lower than 1) along the following steps. \cite{2017A&A...600A.137F} and \cite{2018A&A...619A..47L} illustrated that in AGB simulations (including the one of Table~\ref{simus}) non-stationary convection (e.g., merging down-drafts or other localized events) manifests into giant convective cells, which change topology on very long timescales, and short-lived small surface granules. Moreover, the authors managed to measure the cycles of outward moving shocks and material falling back toward the star and quantify a period for the pulsation modes. 

\begin{figure}
  \centering
    \begin{tabular}{c}
      \includegraphics[width=1.\textwidth]{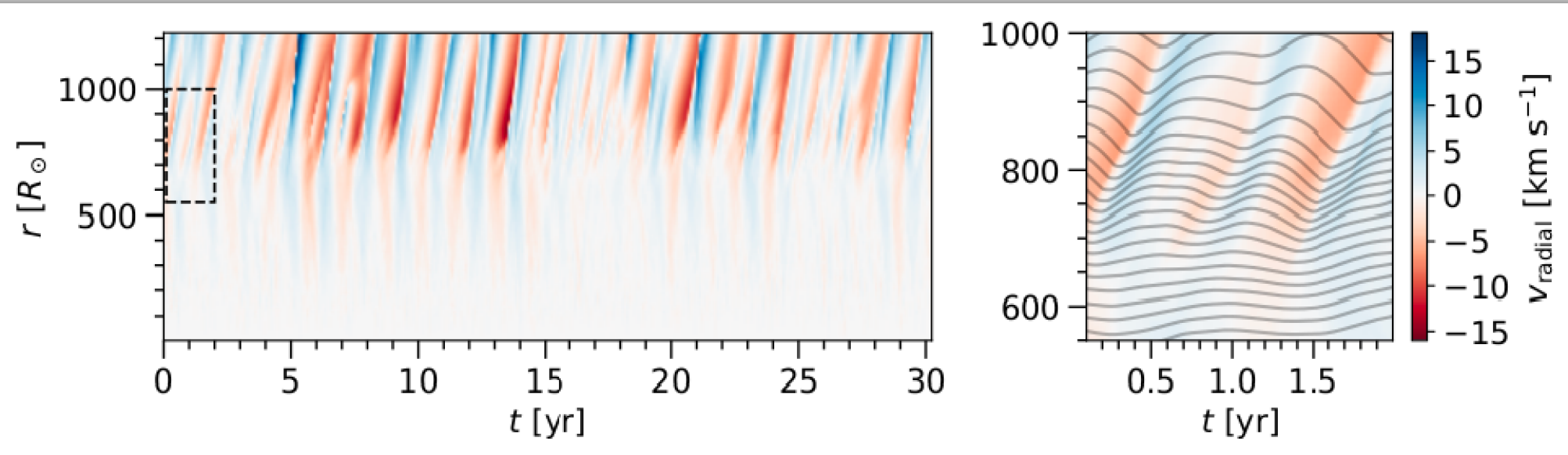}\\
  \includegraphics[width=0.6\hsize]{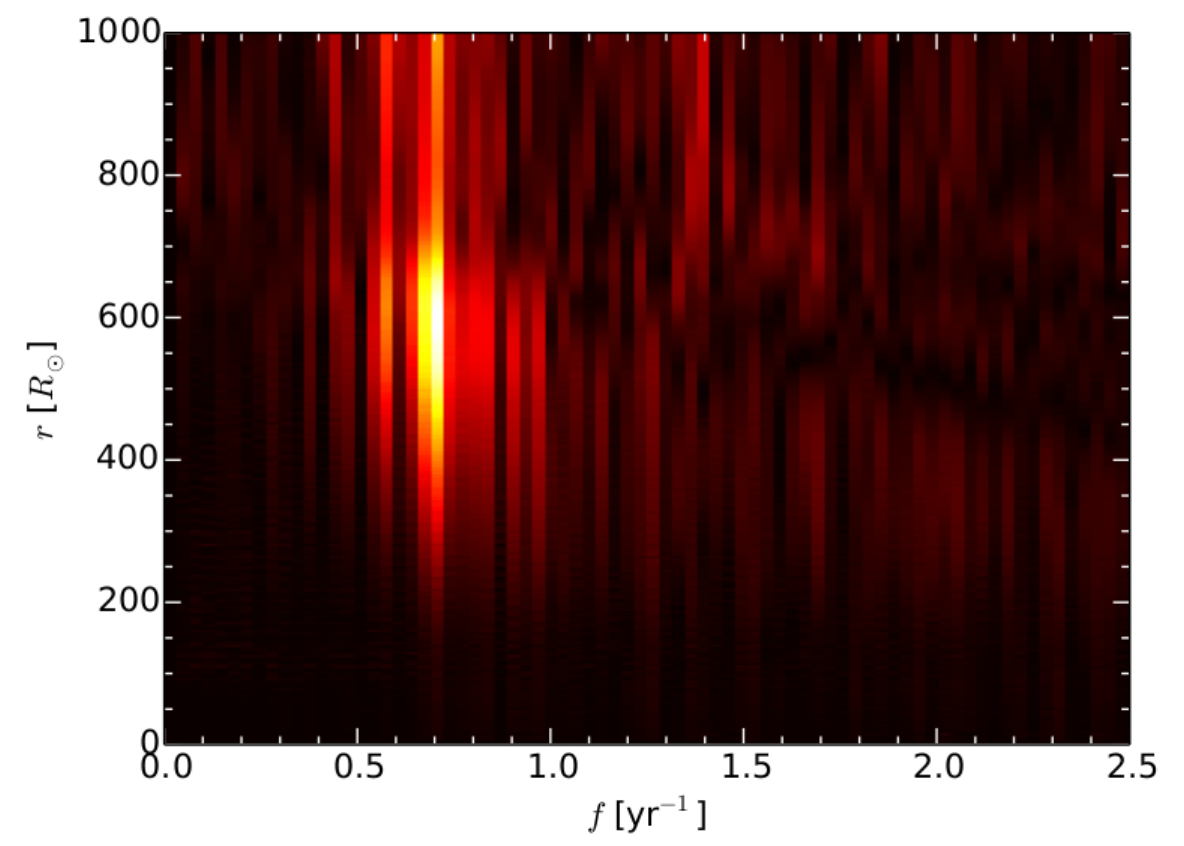}
      \end{tabular}
\caption{\emph{Top panels:} spherically averaged radial velocities for the whole run time and radial extent of the RSG simulation of Table~\ref{simus}. The different colors show the average vertical velocity at that time and radial distance. The rightmost panel displays the area in the dashed rectangle of left panel with mass-shell movements plotted as iso-mass contour lines. The velocity range and color is the same as in Fig.~\ref{fig:radialvel1}. \emph{Bottom panel:} Normalised power spectra derived from the velocity fields from above, mapped over frequency and radial distance.}
\label{fig:radialmean}       
\end{figure}

The radial motions in the atmosphere can then be retrieved using averages over spherical shells of the radial velocities for each snapshot of the simulation (Fig.~\ref{fig:radialmean}, top panels) for the RSG simulation of Table~\ref{simus}. As in \cite{2017A&A...600A.137F} and \cite{2023A&A...669A..49A}, the behavior of the inner part of the model differs from that of the outer layers: below $\sim600\, R_\odot$ (the nominal radius is $616\, R_\odot$) the velocity field is rather regular and coherent over all layers, close to a standing wave. The different slopes visible in the outer layers (above $\sim600\, R_\odot$) are clearly indicating the presence of propagating shock waves but in a less regular and smooth way than in the AGB case (as a matter of comparison, see Fig.~5 of Freytag et al. 2017). The periodic behavior of this movement can be extracted using a Fourier analysis and plotted (Fig.~\ref{fig:radialmean}, bottom panel). A clear prominent frequency is noticeable at $R\sim600\, R_\odot$, while far outside in the shock-dominated atmosphere the possible other modes are significantly smaller in amplitude than in the interior of the star. A Gaussian distribution was fitted in the frequency domain to characterize the largest signal. The resulting period of pulsation is $P_{\mathrm{puls}}=1.456\pm0.209\,\mathrm{ yr}$.\\
These radial pulsations are not transient phenomena introduced by the initial conditions but the outcome of a mode-excitation mechanism caused by the convective motions in the stellar interior, as confirmed for the first time for AGB simulations by \cite{2017A&A...600A.137F}.

\subsubsection{The tomographic method}\label{sectiontomography}

\begin{figure}
  \centering
  \begin{tabular}{c}
      \includegraphics[width=0.6\textwidth]{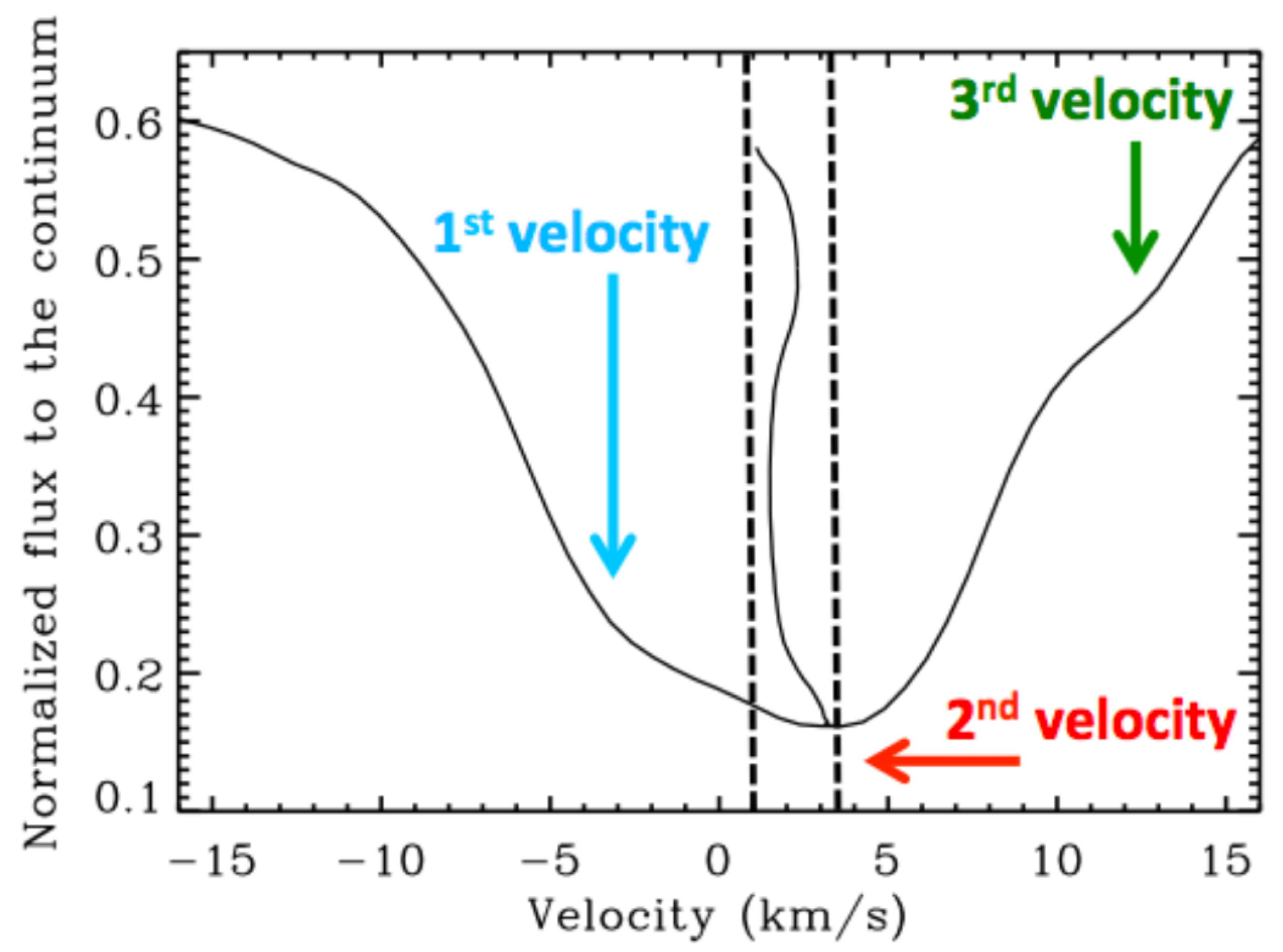}
      \end{tabular}
\caption{Example of a synthetic spectral line of the Ti I at 6261.11~\AA\ for a snapshot of a 3D RSG simulation \citep{2015ASPC..497...11C}. The vertical dashed line shows the spanned velocities of the line bisector. The different arrows and colors display the approximate position of different velocity components, resulting from the heterogeneous velocity field (Fig.~\ref{fig:radialvel1}).}
\label{fig:TiI_line}       
\end{figure}

The study of the velocity field in evolved stellar atmospheres poses a special challenge, as the spectra of these stars are extremely crowded, particularly in the optical domain. Moreover the combined convective and pulsational velocity field have a large impact on the formation of spectral lines producing asymmetries and Doppler shifts (Fig.~\ref{fig:TiI_line}). To disentangle various contributors to the complex line shape, high-resolution spectroscopic measurements coupled with the cross-correlation technique provide a powerful tool to overcome these difficulties. The information relating to the line doubling (i.e. two distinct velocity components, contributing to the line shape) is in fact distributed among a large number of spectral lines, and can be summed up into an average profile, or more precisely into a cross-correlation function (CCF). If the end correlation of the stellar spectrum with a mask involves many lines, it is possible to extract the relevant information from very crowded and/or low signal-to-noise spectra. Moreover, radial velocities must be traced on both spatial and temporal scales to allow detailed reconstruction of atmospheric motions in evolved stars. A recently established tomographic method is an ideal technique for this purpose. It must be stated right away that the word ”tomography” is used here with a meaning close to its etymological roots ("slice, section"). 

The tomographic method \citep{2001A&A...379..288A} aims to recover the temporarily-resolved line-of-sight velocity distribution as a function of depth in the stellar atmosphere. The technique is based on sorting spectral lines according to their formation depths. The atmosphere is split into different slices corresponding to specific optical depth ranges. For each slice, a spectral mask is then constructed, which contains the wavelengths of lines forming in the corresponding range of optical depths. The cross-correlation functions (CCFs) of the masks with stellar spectra reflects the average shape and radial velocity (RV) of lines forming in a given optical depth range.

\begin{figure}
   \centering
   \includegraphics[width=0.9\textwidth]{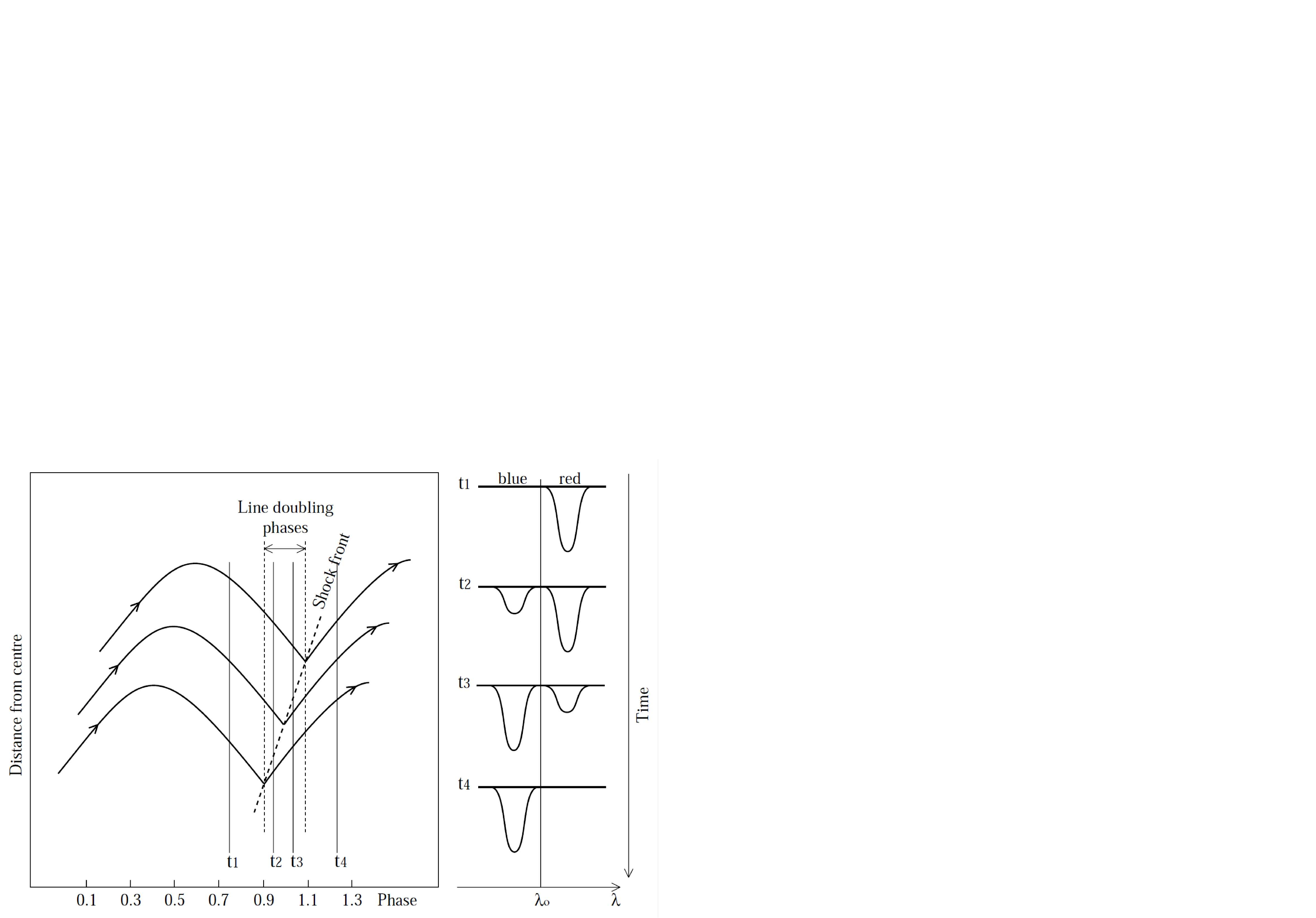}
      \caption{Illustration of the Schwarzschild scenario: temporal sequence following the intensity of the red and blue components of absorption lines, when a shock wave propagates through the photosphere. Figure from \citet{2000A&A...362..655A}.}
         \label{Fig:schwarz_scenario}
   \end{figure}

\begin{figure}
   \centering
   \includegraphics[width=0.9\textwidth,angle=180]{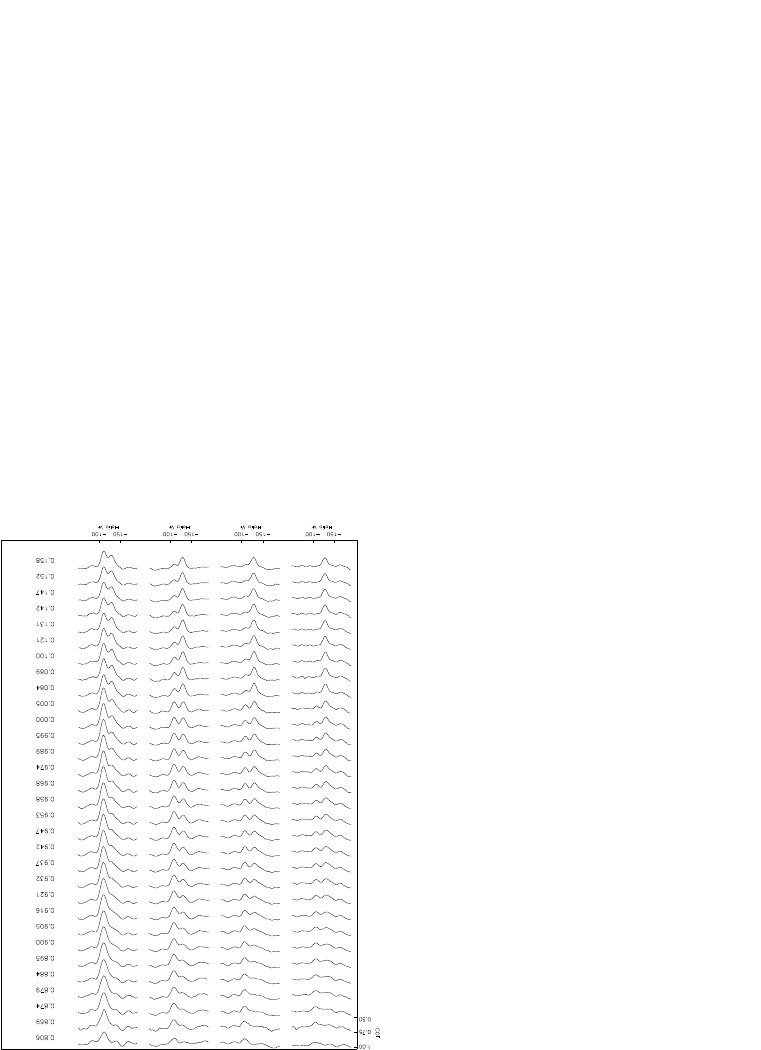}
      \caption{Sequence of cross-correlation profiles of the Mira star RT Cyg obtained with four tomographic masks, around maximum light. Columns correspond to different masks, the leftmost mask probes the innermost atmospheric layer. Rows correspond to variability phases varying from 0. to 1. The cross-correlation profiles follow the Schwarzschild scenario: a single red component is progressively transforming into a single blue component. Reproduced from \citet{2016ASSL..439..137J}.
            }
         \label{fig:shockwave}
   \end{figure}

The concept of the tomographic method originates from the recognition by \citet{schwarzschild1954} that the doubling of the H-$\alpha$ line observed in Cepheid variables of the W~Vir type follows a specific time sequence that may be explained by a shock front passing through the line-formation zone. Line doubling has since been observed in many other lines, both in the infrared and in the optical. For long-period variables (LPVs) of Mira type, \citet{1955PASP...67..199M} was the first to suggest that the observed spectral changes may be explained by the presence of a shock wave moving outwards. In pulsating atmospheres of LPVs or Cepheids, what has since been known as the Schwarzschild scenario \citep[][see Fig.~\ref{Fig:schwarz_scenario}]{schwarzschild1954} states that, when the shock front is located far below the line-forming region, the stellar atmosphere contains only falling material so that the spectral lines are red-shifted. Then, as the shock wave lift up, a blue component appears and strengthens. This outward motion of the shock front has been revealed by the tomographic method applied to Mira variables (Fig.~\ref{fig:shockwave}). One example is reported in \cite{2000A&A...362..655A} for a particular AGB stars, namely Mira, for which the tomographic method clearly unveils the Schwarzschild scenario as a line-doubling (ie., the shock signature, Fig.~\ref{Fig:schwarz_scenario}) appearing in diverse layers of time-consecutive CCF probing, from below to  above the stellar surface. In other words, the tomographic method revealed the upward motion of the shock fronts combining temporal and spatial variations of the CCFs \citep{2016ASSL..439..137J}.

\begin{figure}
   \centering
   \includegraphics[width=1.\textwidth]{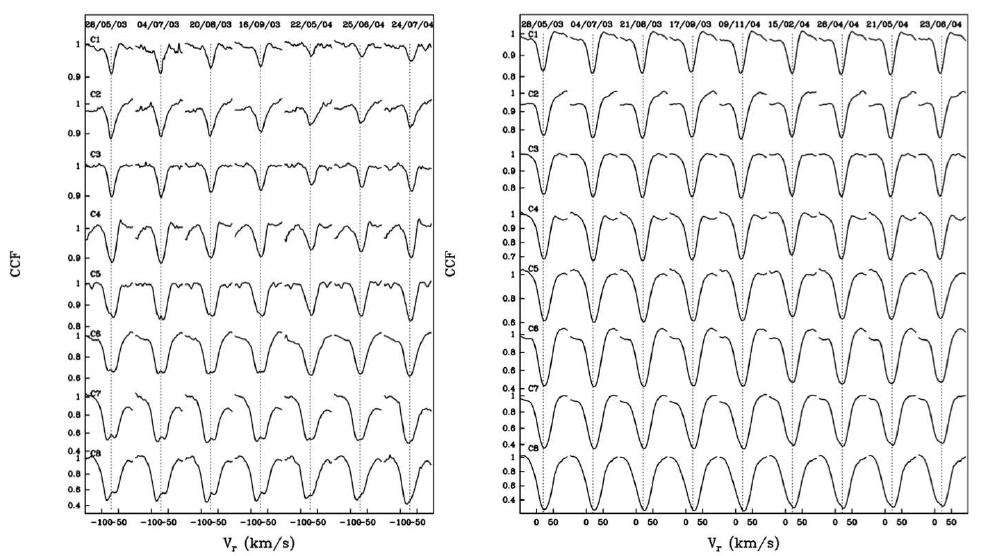}
      \caption{Sequence of cross-correlation profiles of the RSG stars SW Cep (left) and $\mu$~Cep (right) obtained for eight tomographic masks.  Rows correspond to different masks, mask C1 probes the innermost atmospheric layer. Columns correspond to different observing epochs. In contrast to Mira stars, the CCFs of RSGs do not follow the Schwarzschild scenario. Figure from \citet{2007A&A...469..671J}.
      }
         \label{Fig:RSG_Josselin}
   \end{figure}

The tomographic method of \citet{2001A&A...379..288A} was applied to short time-series
spectra of RSG stars by \citet{2007A&A...469..671J}. This study detected line asymmetries
in resulting cross-correlation profiles and the absence of the Schwarzschild scenario (Fig.~\ref{Fig:RSG_Josselin}). Thus, RSGs are interesting targets for a more detailed tomographic study.

A detailed description of the tomographic technique and its recent developments can be found in \citet{2018A&A...610A..29K}. The method relies on the design of spectral templates (so-called masks) containing lines forming at given, pre-specified ranges of optical depths. The original implementation of the method \citep{2001A&A...379..288A} relied on the Eddington-Barbier approximation, stating that in a grey atmosphere, the flux emerges from the layer located at optical depth $\tau_{\lambda} = 2/3$. Thus a line at wavelength $\lambda$ was supposed to form at optical depth $\tau_{\lambda} = 2/3$. An appropriate model atmosphere is then used to convert this monochromatic optical depth $\tau_{\lambda} = 2/3$ into a reference optical depth (at 5000~$\AA$ for instance), which is then a proxy for geometrical depth. As shown by \citet{1986A&A...163..135M}, the Eddington-Barbier approximation gives however a valid depth of line formation only for sufﬁciently strong lines. In the recent implementation of the tomographic method \citep{2018A&A...610A..29K}, the contribution function to the line depression is used in order to correctly assess the depth of formation of spectral lines. 
\citet{1996MNRAS.278..337A} derived the contribution function\footnote{The amount of light contributed to the surface flux as a function of depth in the atmosphere} (CF) for a single ray (SR) of a 1D plane-parallel atmosphere:

\begin{equation}
  {\rm CF_{SR}}(\log \tau_0,{\lambda}) = (\ln10) \, \mu^{-1} \frac{\tau_0}{\kappa_{c,0}} \kappa_{l_{\lambda}} \,(I_{c_{\lambda}} - S_{l_{\lambda}}) \, e^{-\tau_{\lambda}/\mu}    
\label{1DCF}
\end{equation}
\ 
and for the whole stellar disk (DI): 

\begin{equation}
{\rm CF_{DI}}(\log \tau_0,{\lambda}) = {2\pi}\,(\ln10) \frac{\tau_0}{\kappa_{c,0}} \int_0^1 \kappa_{l_{\lambda}} \,(I_{c_{\lambda}} - S_{l_{\lambda}}) \, e^{-\tau_{\lambda}/{\mu}} \, d\mu 
\label{Eq:cfalbrow}
\end{equation}
\
In the above equations, $\tau_0 = \int \kappa_{c,0} \, \rho \, {\rm d}x$ is the continuum optical depth at a reference wavelength $\lambda_0 = 5000$~\AA, $\rho$ is the density, $\kappa_{c,0}$ is the continuum absorption coefficient at $\lambda_0$, $I_c$ is the continuous intensity, $I_l$ is the line intensity, $\kappa_l$ is the line absorption coefficient, $S_l$ is the line source function, $\tau_{\lambda} = \int \kappa_{\lambda} \, \rho \, {\rm d}x$ is the optical depth along the ray, $\kappa_{\lambda}$ is the total absorption coefficient at wavelength $\lambda$, $\mu=\cos{\theta}$ is the cosine of the angle $\theta$ between the line of sight and the radial direction ($\mu = 1$ for the direction towards the observer). The $\tau_0$ scale can be used as a proxy to the geometrical depth $x$.

\begin{figure}
  \centering
  \begin{tabular}{c}
      \includegraphics[width=0.9\textwidth]{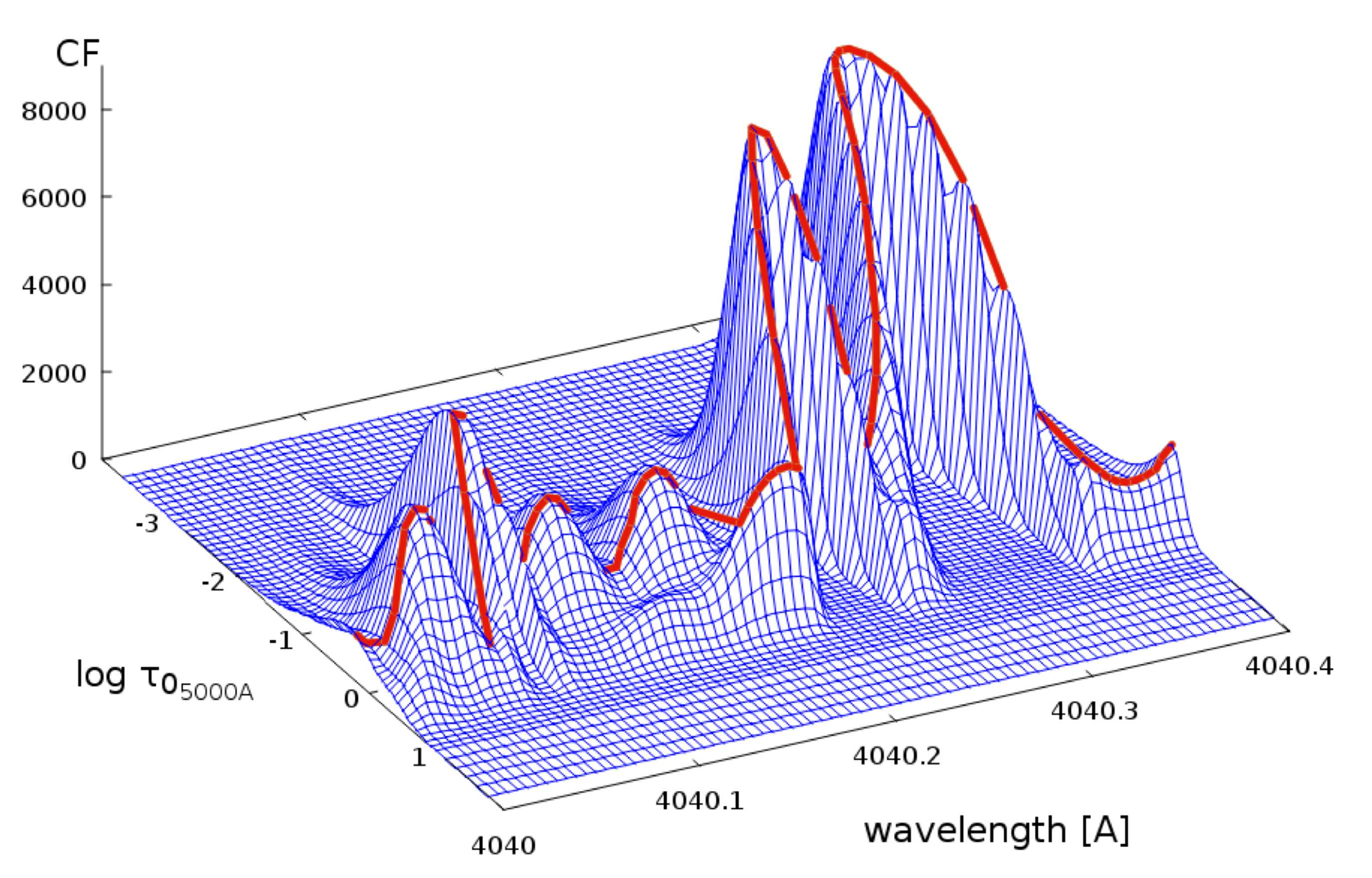}
      \end{tabular}
\caption{An example of the Contribution Function emerging from a 1D hydrostatic model of a RSG star \citep[figure from][]{2018A&A...610A..29K}.  The crest line shown in red defines the depth function illustrated in Fig.~\ref{fig:tomoschema1}.}
\label{fig:1DCF_RSG}      
\end{figure}

\begin{figure}
  \centering
    \begin{tabular}{cc}
      \includegraphics[width=0.48\hsize]{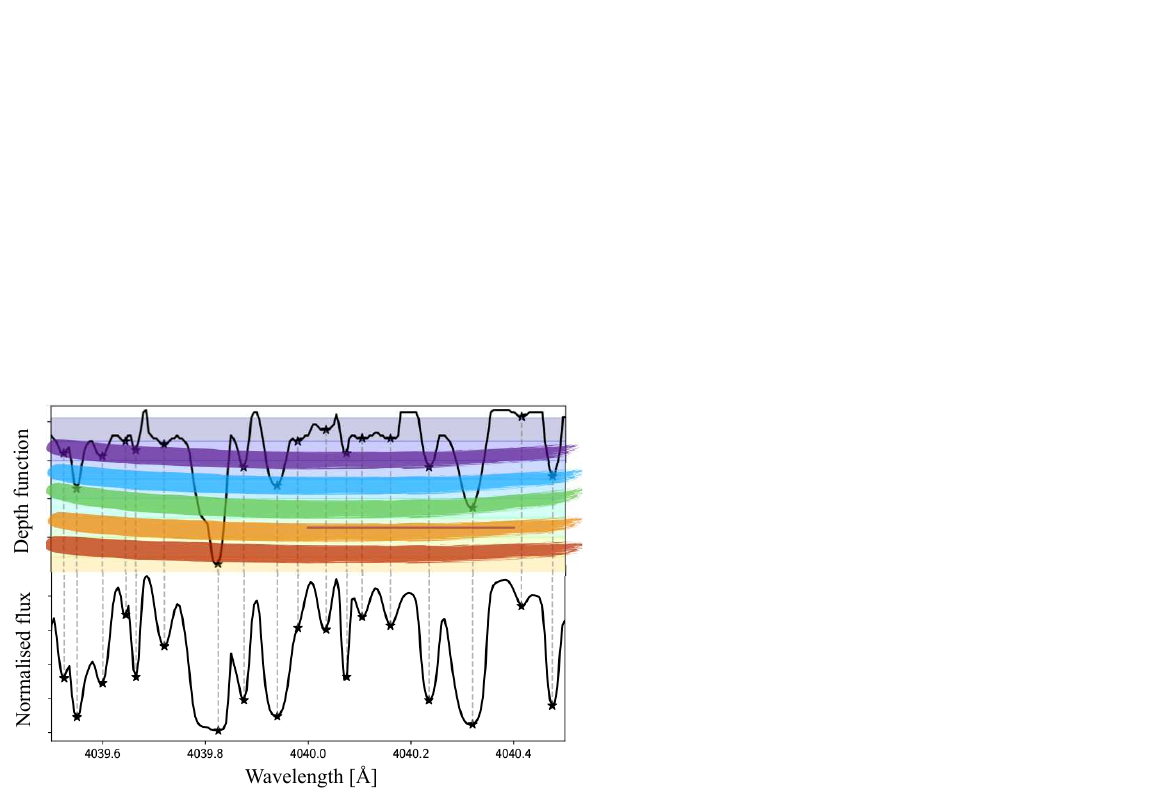}
           \includegraphics[width=0.5\hsize]{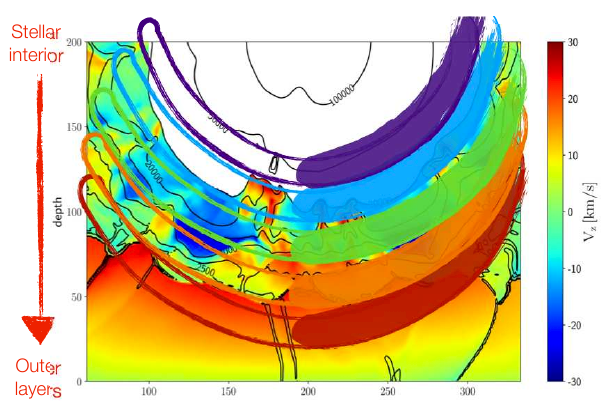}
      \end{tabular}
\caption{Sketch of the mask construction process. \emph{Left panel: } The depth function (top) and the normalised synthetic flux (bottom) are shown as a function of wavelength for a 1D RSG model. The depth function corresponds to the crest line of the CF shown in Fig.~\ref{fig:1DCF_RSG}. The comparison of the depth function with the synthetic spectrum shows that the cores of spectral lines form in outer layers while the wings form in deeper layers. The different colors sketch the approximate position of the masks used to divide the atmosphere. \emph{Right panel:} The different atmospheric layers spanned by the masks are sketched over a slice of a RSG RHD simulation representing the line-of-sight velocity ($V_z$). Note that the masks are not assumed to be concentric spherical shells.}
\label{fig:tomoschema1}      
\end{figure}

\begin{figure}
  \centering
      \includegraphics[width=0.9\hsize]{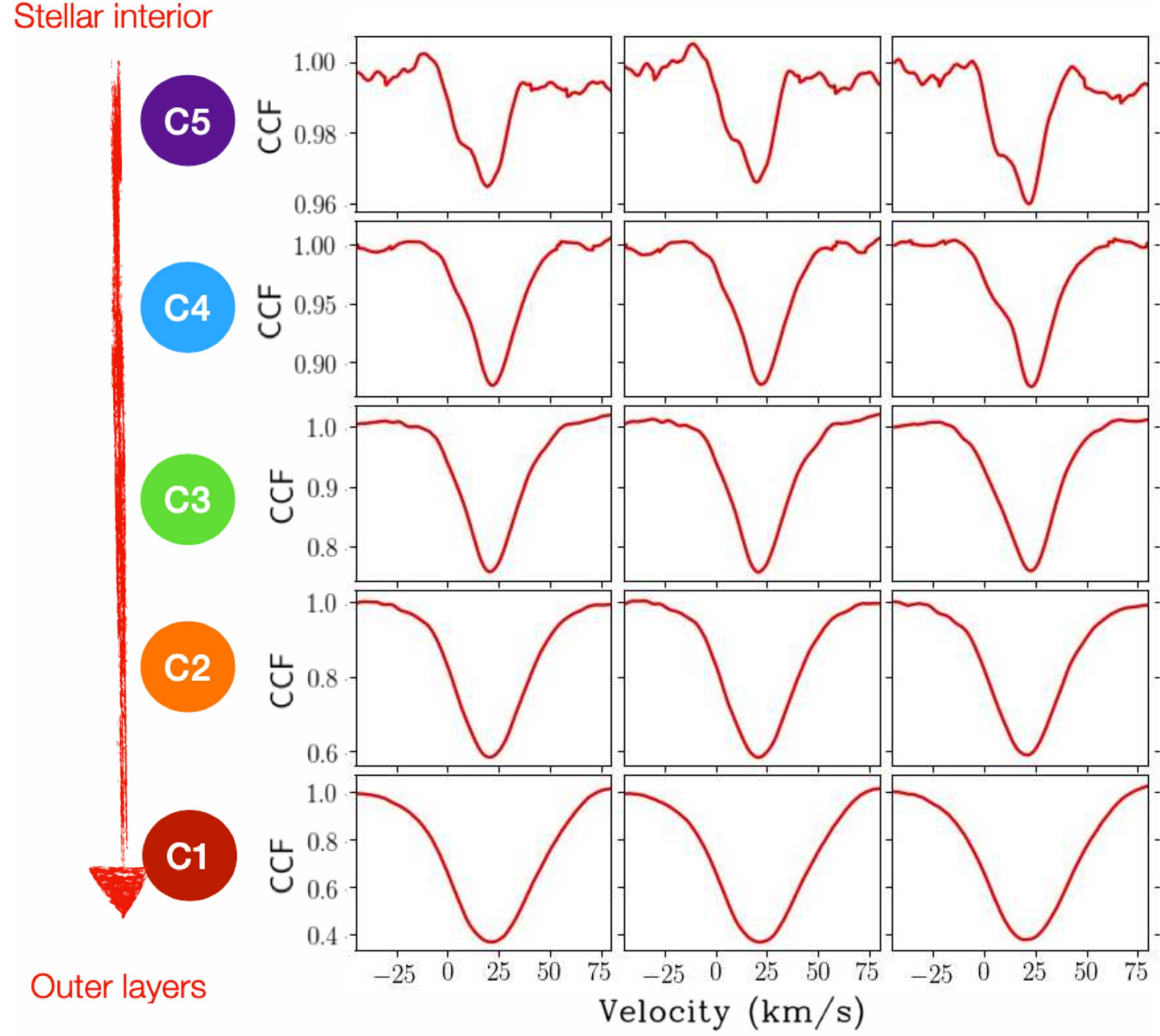}
\caption{Example of a temporal sequence (columns) of cross correlation functions (CCF, obtained by applying a set of five tomographic masks to a series of observed stellar spectra. The masks (C1 to C5) have the same colors as in Fig.~\ref{fig:tomoschema1}.}
\label{fig:tomoschema2}      
\end{figure}

\citet{2018A&A...610A..29K} implemented the computation of the CF in the 1D radiative transfer code TURBOSPECTRUM \citep{2012ascl.soft05004P}, which performs spectrum synthesis from static one-dimensional (1D) MARCS model atmospheres \citep{2008A&A...486..951G}. Fig.~\ref{fig:1DCF_RSG} illustrates an example of the CF emerging from a 1D hydrostatic model of RSG star. The set of $\tau_0$ corresponding to the maximum CF for each wavelength defines a crest line called "depth function". The depth function resembles the shape of a spectrum but instead of intensity (or flux) it indicates the central optical depths $\tau_0$ at which the emergent intensity at wavelengths $\lambda$ forms (left panel of Fig.~\ref{fig:tomoschema1}). This allows to split the atmosphere into different optical-depth layers (colored bands in Fig.~\ref{fig:tomoschema1}) and construct tomographic masks across the stellar atmosphere. Each mask contains the central wavelengths of spectral lines whose depth function minimum falls within a range of optical depths defined by a given layer. The cross-correlation of the masks with observed or synthetic stellar spectra provides the distribution of the line-of-sight velocity field inside a given atmospheric slice (Fig.~\ref{fig:tomoschema2}). 

\subsubsection{Recovering velocity fields across the atmosphere in RSG stars}\label{section_hysteresis}

\begin{figure}
  \centering
  \begin{tabular}{c}
      \includegraphics[width=0.7\textwidth]{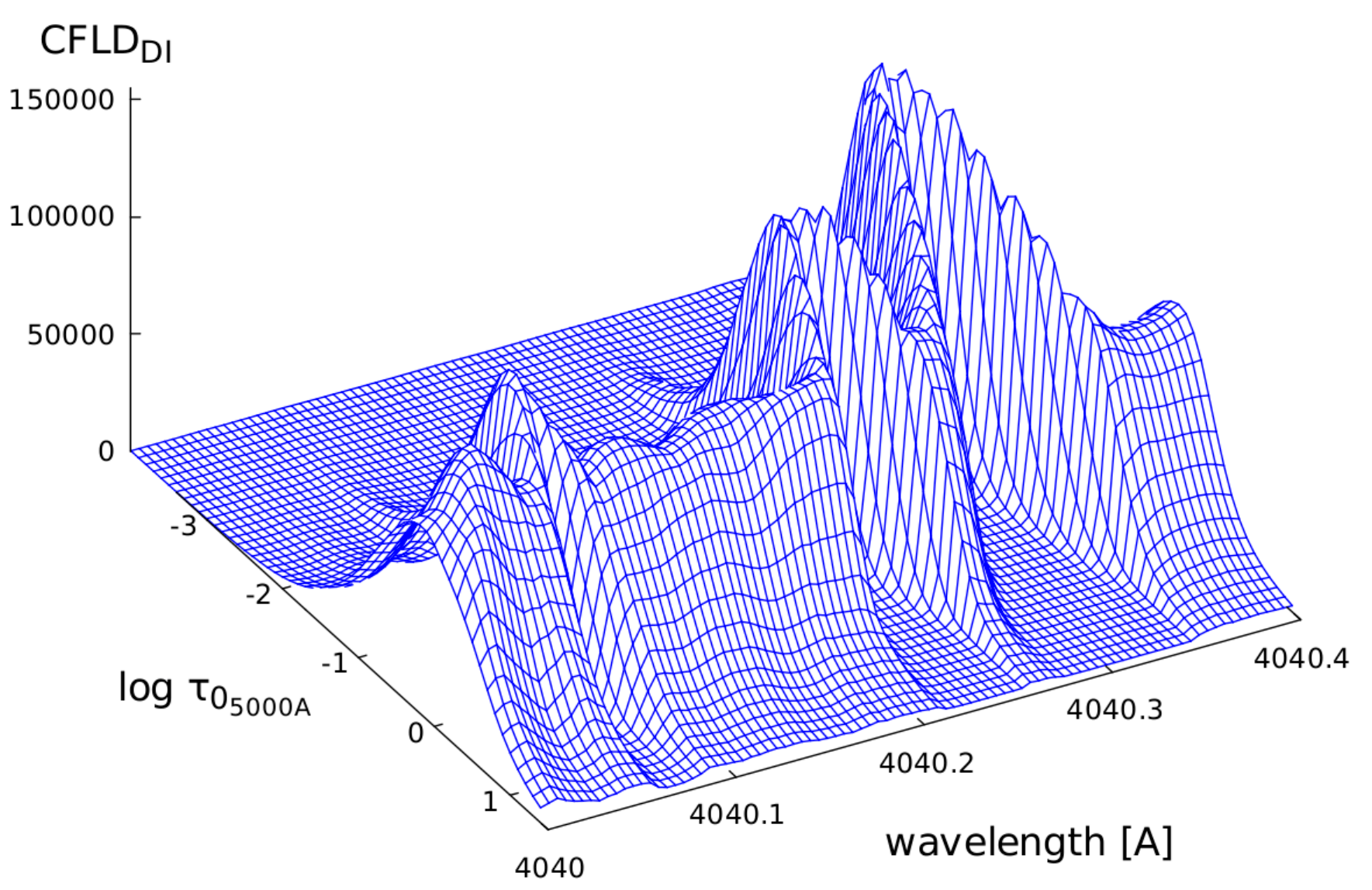}
      \end{tabular}
\caption{An example of the 3D CF for a 3D RHD simulation of a RSG star from Table~\ref{simus} for the same wavelength range as Fig.~\ref{fig:1DCF_RSG} \citep[figure from][]{2018A&A...610A..29K}.}
\label{fig:3DCF_RSG}      
\end{figure}

The validation of the tomographic method on 3D RHD simulations needs the post-processing calculations of high spectral resolution synthetic spectra from the 3D thermodynamical structures. For this purpose, a possible option is the use of the 3D pure-LTE radiative transfer code Optim3D \citep{2009A&A...506.1351C}. This code computes synthetic spectra and intensity maps for any snapshots of RHD simulations and 
uses pre-tabulated extinction coefficients generated with the MARCS code \citep{2008A&A...486..951G} and by adopting the solar composition of, e.g., \cite{2009ARA&A..47..481A}. Then, the extinction coefficients are interpolated at each Doppler-shifted wavelength in each cell along the ray. The numerical integration of the equation of radiative transfer gives the intensity emerging towards the observer at that wavelength and position. This calculation is performed for every $x$-/$y$-point of a face of the simulation cube, and for all the required wavelengths. Using these spectra, the tomographic method has been tested by recovering the velocity field distribution across the atmosphere of a 3D RHD simulation of an RSG star in \citet{2018A&A...610A..29K}. The authors derived the 3D CF for the "star-in-the-box" simulation, which was implemented in the radiative transfer code Optim3D:

\begin{equation}
CF_{\rm 3D} (\tau_0) = {\rm ln} 10 \sum_{i,j=0}^{N} \, \tau_0 \, \frac{\kappa_{l,i,j}}{\kappa_{0,i,j}} \, \bigg[I_{c,z} \Big(x_i,y_j,z(\tau_0)\Big) - \\
S_{l,z} \Big(x_i,y_j,z(\tau_0)\Big)\bigg] e^{-\tau } \frac{ \Delta x_i \Delta y_j}{R^2}
\label{Eq: finalCU}
\end{equation}
\
where $N$ is the number of grid points in the numerical box along the line-of-sight $z$-axis, $i$ and $j$ are ray numbers along the x and y axes. For the equidistant grid $\Delta x_i = \Delta y_j$. All quantities in the equation depend on wavelength $\lambda$. Fig.~\ref{fig:3DCF_RSG} illustrates an example of the 3D CF computed with Eq.~\ref{Eq: finalCU} for the same wavelength range as the 1D CF from Fig.~\ref{fig:1DCF_RSG}. Similar to the 1D CF, the 3D CF is characterized by a single maximum in the $\log \tau_0$ direction. 

\begin{figure}
  \centering
    \begin{tabular}{cc}
      \includegraphics[width=0.48\hsize]{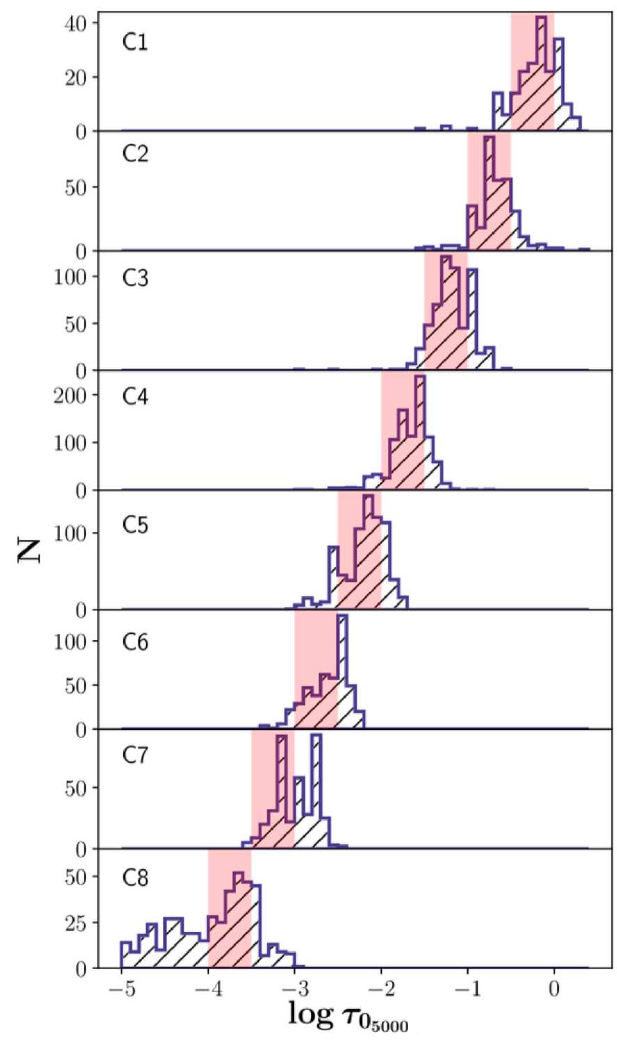}
      \hspace{1cm}
           \includegraphics[width=0.35\hsize]{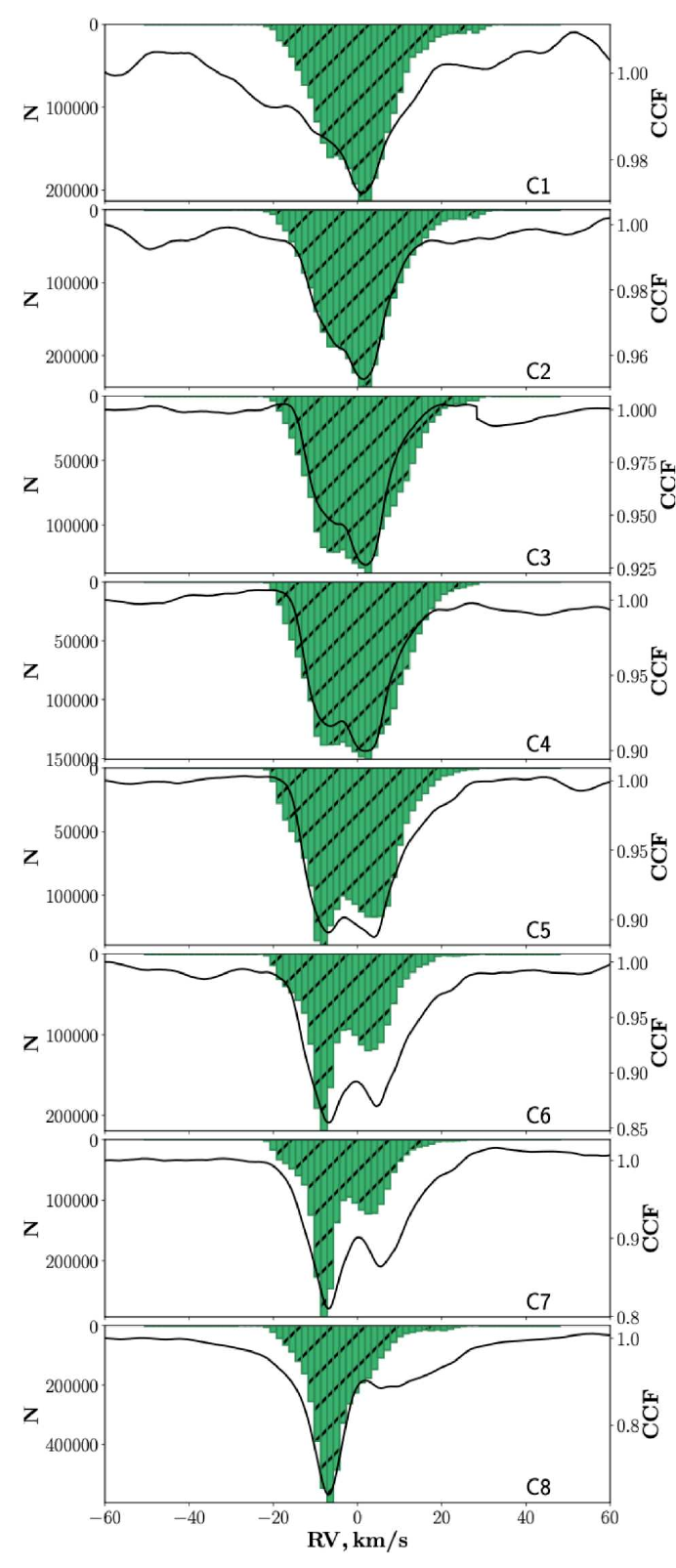}
      \end{tabular}
\caption{\textit{Left panel:} Distribution of the line-formation depths in a 3D RSG snapshot (blue color) for a set of eight tomographic masks. The red bands represent the optical depth ranges used for selecting lines in a given mask from the 1D contribution function. \textit{Right panel:} Reconstruction of the velocity field in the 3D snapshot as a function of the optical depth in the atmosphere. Adopted from \citet{2018A&A...610A..29K}.}
\label{fig:tomo_validation}     
\end{figure}

To test the tomographic method, a set of eight tomographic masks is constructed using a 1D hydrostatic MARCS model of a RSG star with parameters matching those of the 3D RHD simulation. The left panel of Fig.~\ref{fig:tomo_validation} compares the distributions of the line-formation depths (represented by the maxima of the CF in the $\log \tau_0 - \rm CF$ plane) in different masks for the 3D snapshot and the 1D model used for the masks construction. The 1D and 3D line-formation depths qualitatively overlap in all masks from the inner (mask C1) to the outer layers (mask C8). Thus, the constructed masks from a 1D model are usable to probe different atmospheric depths in a dynamical atmosphere. 

As the next step, the tomographic masks are cross-correlated with a synthetic spectrum computed from a 3D snapshot of the same simulation used above.
The resulting cross-correlation functions (CCFs) are shown in the right panel of Fig.~\ref{fig:tomo_validation} together with the histograms of the line-of-sight ($V_z$) velocities extract from the 3D structure of the snapshot. The velocity field overlaps the CCF functions across the atmosphere and, as a consequence, the line-of-sight velocity distribution can be fully recovered. In conclusion, the emerging spectral lines spread over different optical depths due to the non-radial convective movements characterizing the stellar atmosphere. This theoretical validation of the tomographic method opens a new window for the study of stellar dynamical cycles in evolved stars, and in particular RSGs.

\begin{figure}
\centering
\includegraphics[width=0.99\linewidth]{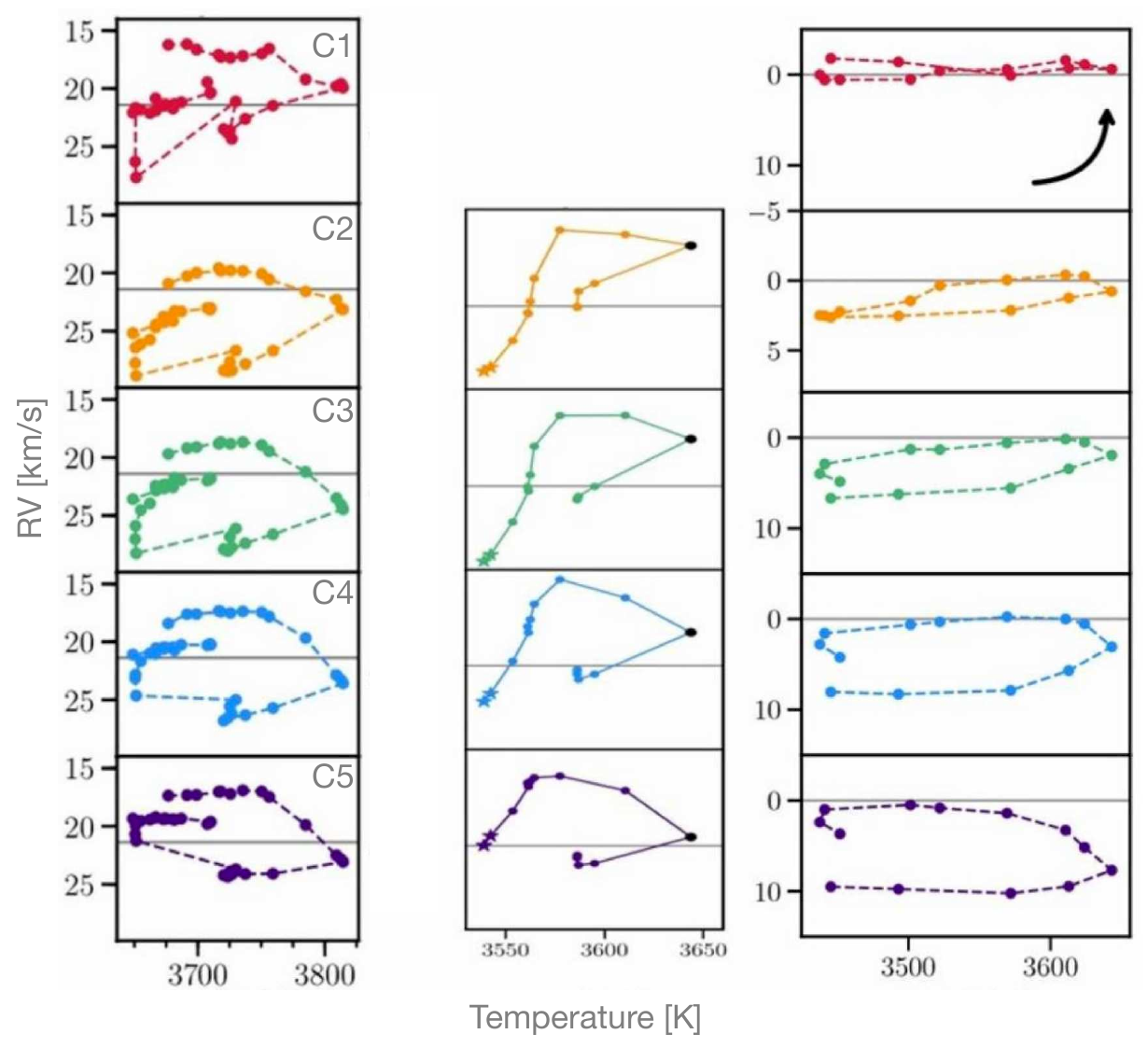}
 \caption{Hysteresis loops between the temperature and RV variations for a single light cycle of $\mu$~Cep \citep[left, from][]{2019A&A...632A..28K}, Betelgeuse \citep[middle, from][]{2021A&A...650L..17K} and the 3D simulation \citep[right, from][]{2019A&A...632A..28K}, and for the different tomographic masks (rows). The arrow indicates the direction of evolution along the hysteresis loops.}
   \label{fig:obs_cycle}
\end{figure}

\begin{figure}
\centering
\includegraphics[width=1.\hsize]{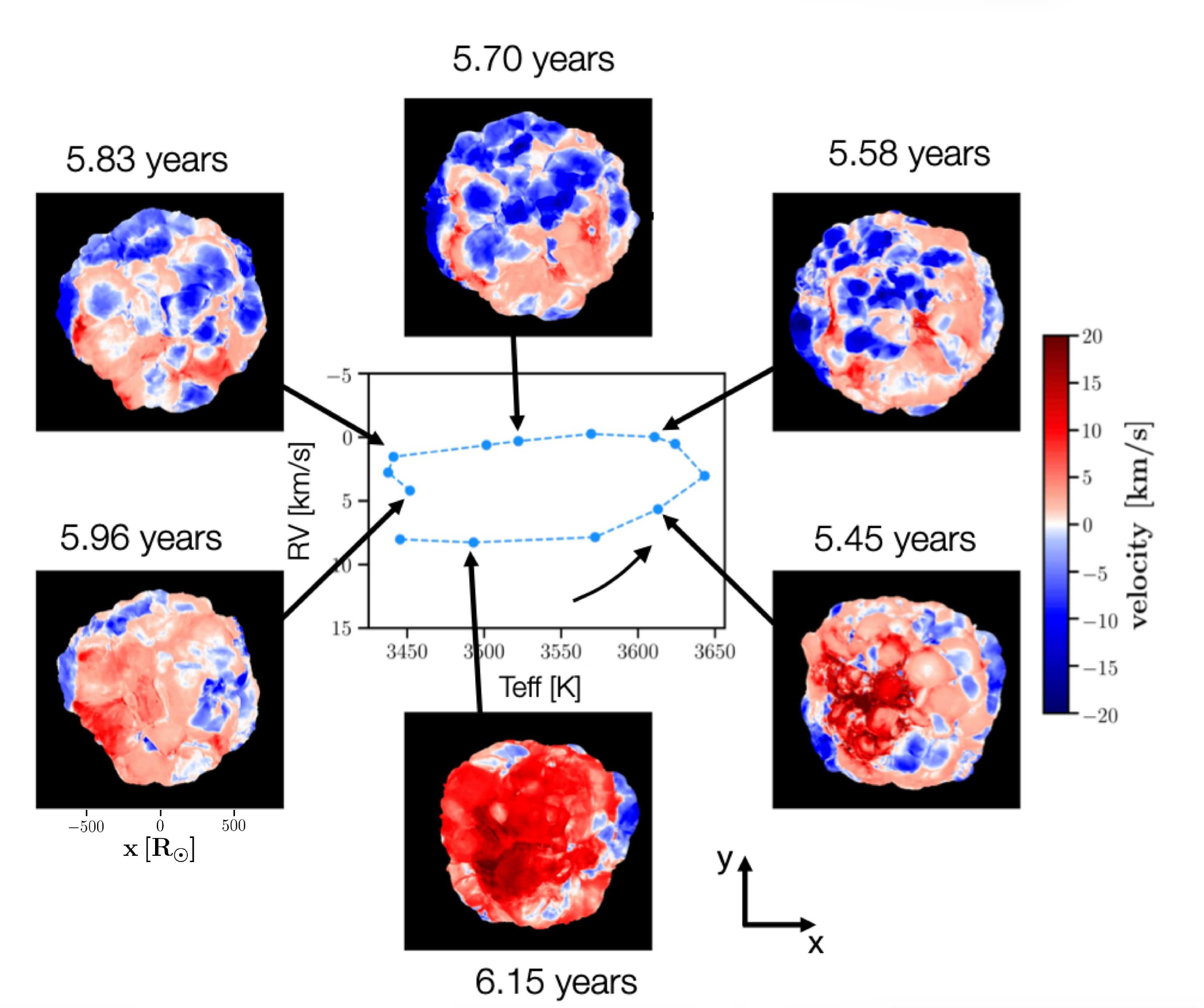}
 \caption{Velocity maps for the C4 mask and for different 3D snapshots along the hysteresis loop for that mask ( right panel of Fig.~\ref{fig:obs_cycle}). The red and blue colors correspond to inward and outward moving material, respectively. The central panel shows the behaviour of the radial velocity. The arrow indicates the direction of the evolution along the hysteresis loop, characteristic signature of the convective turn-over of the material in the stellar atmosphere.}
   \label{fig:cycle}
\end{figure}

\subsubsection{Convective cycles in RSG stars: the cases of $\mu$~Cep and Betelgeuse}

The first application of the tomographic method to interpret stellar dynamical cycles in RSGs is reported by \citet{2019A&A...632A..28K} and \citet{2021A&A...650L..17K} for two prototypical and extensively observed RSG stars: $\mu$~Cep and Betelgeuse. In particular, these studies aims at explaining the short-period, photometric variability of RSGs, whose nature is still debated due to the semi- and irregular character of the variations. 

To identify the acting mechanism(s), the tomographic method is applied to several years of regular time-series of high-resolution (resolving power of $\lambda/\Delta\lambda\sim$80000) optical spectra (3800$-$9000 \AA) of these objects, where a phase shift was detected between the radial velocity and photometric variations \citep{2019A&A...632A..28K,2021A&A...650L..17K}. \\
This phase shift results in a hysteresis loop \citep{2008AJ....135.1450G} in the temperature-velocity plane (left and middle panels of Fig.~\ref{fig:obs_cycle}) with timescales similar to the photometric ones. 

On the 3D RHD simulation side, its atmosphere reveals qualitatively similar hysteresis-like behavior in the temperature--velocity plane (right panel of Fig.~\ref{fig:obs_cycle}) with (quasi-) periods consistent with the observed photometric timescales. In the simulation, the hysteresis process is the characteristic signature of the convective turn-over of the material in the stellar atmosphere. \\
As an example, Figure~\ref{fig:cycle} illustrates
the velocity maps for mask C4, for different 3D snapshots along the hysteresis loop from the right panel of Fig.~\ref{fig:obs_cycle}. They reveal upward (blue) and downward (red) motions of matter extending over large portions of the stellar surface: the relative fraction of upward and downward motions is what distinguishes the upper from the lower part of the hysteresis loop, its top part (zero velocity) being characterized by equal surfaces of rising and falling material. The bottom part of the hysteresis loop occurs, as expected when the stellar surface is covered mostly by down-falling material. Moreover, the sound-crossing timescale in the outer layers is of the same order as the hysteresis-loop timescale. This suggests that hysteresis loops are linked to acoustic waves originating from the turbulent convective flow below the visible atmosphere and propagating upward where they modulate the convective energy flux with shocks.\\
Another example of the use of this method concerns 
the RSG Betelgeuse, which experienced a historic dimming of its visible brightness (whose usual apparent magnitude comprises between 0.1 and 1.0) down to 1.614$\pm$ 0.008 magnitudes around 07-13 February 2020 \citep{2020ATel13512....1G}. Using this tomography method and 3D simulations, \cite{2021A&A...650L..17K} showed the presence of two successive shocks in February 2018 and January 2019 that produced a rapid expansion of part of the atmosphere of Betelgeuse. The authors provided an explanation of the phenomenon arguing that the sudden increase in molecular opacity in the cooler upper atmosphere of Betelgeuse should have caused obscuration of the star and the observed unusual decrease of its brightness. Other explanations based on the circumstellar environment where dust grains form are also possible 
(e.g. \citealt{2021Natur.594..365M}). For more details, the reader can refer to the recent review of \cite{2023A&G....64.3.11W}.


\subsection{Athena++ applications: Calibrating Convection for 1D models and Predicting Supernova Signatures}\label{sec_AthenaResults}

The Athena++ simulations have thus far been applied primarily in the context of 1D stellar and supernova modeling -- in particular, stellar convection and supernova shock breakout. We discuss these applications in the following part of this Section.

\subsubsection{Calibrating Mixing Length Theory from the Deep Stellar Interior}

\begin{figure}
\centering
\includegraphics[width=1.\textwidth]{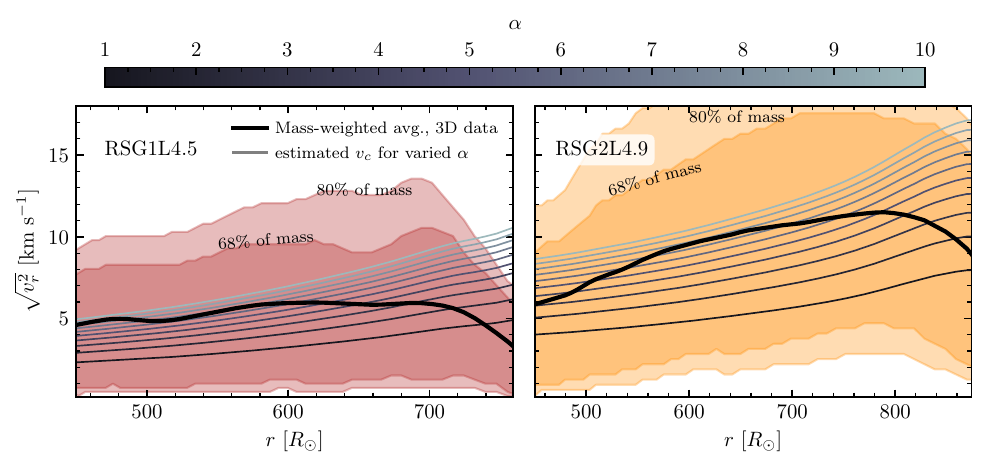}
\caption{\label{fig:alphavels} Radial fluid velocity magnitudes at characteristic snapshots of two Athena++ RSG models compared to MLT (RSG1L4.5, day 4707, left; RSG2L4.9, day 4927, right). Mass-weighted average velocities are shown as thick black lines, with 68\% and 80\% of the mass lying within the dark and light shaded regions, respectively. Greyscale lines indicate convective velocities predicted from MLT given the volume-averaged temperature and density profiles and the model luminosity, for integer values of $\alpha=$1---10 where higher values of $\alpha$ predict larger convective velocities. The plots are truncated where the turbulent motions do not resemble MLT-like convection, which lies inside the region where $\nabla\langle s\rangle\le0$. Figure from \citet{Goldberg2022a}.} 
\end{figure}

Convection is a fundamentally 3D process, yet it is a dominant mode of energy transport and compositional mixing in stellar interiors and envelopes, necessitating prescriptions for its effects on 1D stellar evolution models. The Mixing Length Theory (MLT), first formulated for stellar interiors by \citet{BohmVitense1958} and expanded by \citet{Henyey1965} to model the transition from convective to radiative transport, serves as the default prescription for convection in the vast majority of 1D stellar evolution codes \citep[see][for a detailed review]{Joyce2023}. 

A key element of the MLT prescription is that the characteristic length scale of transport which determines the efficiency of convection, i.e. the mixing length $\ell$, taken locally to be a free parameter $\alpha$ times the pressure scale height, $\ell=\alpha H$. Especially if the \emph{envelope} is convective, assumptions about $\alpha$ and other MLT parameters strongly influence the stellar radii and $T_{\rm eff}$ output by 1D codes \citep[e.g.][]{Stothers1995,Massey2003,Meynet1997,Meynet2015,Joyce2023} (in addition to
assumptions relevant to the structure and location of internal convective boundaries). Such values of $\alpha$ are often observationally calibrated by comparing stellar models to the Sun \citep[e.g.][and many others]{Charbonnel1993,Cinquegrana2022} and other well-characterized stars \citep[e.g.][]{Guenther2000,Chun2018,Joyce2018}. 

Computational RHD models of convection enable a theoretical calibration of MLT free parameters. Early 2D RHD simulations \citep{Ludwig1999} and 3D
simulations by \citet{Sonoi2019} calibrated MLT in $L\ll L_\mathrm{Edd}$ stars, constraining $\alpha\approx1-2$ across a grid of low-luminosity cool giant atmospheres (Red Giants with $T_{\rm eff}>4000$K), in agreement with 
the observational constraints. Other works (e.g. 
\citealt{2014MNRAS.445.4366T,Magic2015,Salaris2015}) recovered similar calibrations in similar $L\ll L_\mathrm{Edd}$ stars, with some trends towards higher efficiency (i.e. higher $\alpha$) in more luminous stars. As convection becomes even more vigorous in the coolest, most luminous stars, such as AGB and RSG stars, individual convective plumes take up larger and larger fractions of the star and global or semi-global simulations in full-star geometry are required, and in many cases convective motions particularly at the stellar surface begin to deviate significantly from MLT \citep[see, e.g., the discussion in][]{2013ApJ...769...18T}.  

\begin{figure}
\centering
\includegraphics[width=0.48\textwidth]{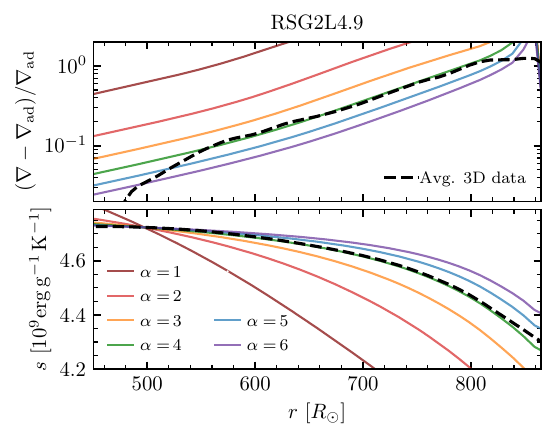}
\includegraphics[width=0.48\textwidth]{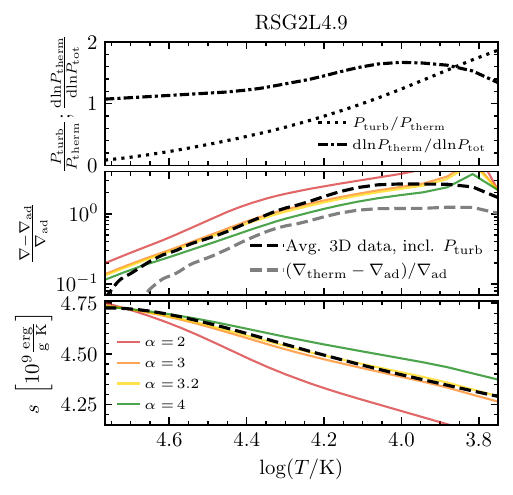}
\caption{\label{fig:MLTcal} Left: Comparison of superadiabaticity (upper left) and specific entropy (lower left) derived from the RSG2L4.9 Athena++ RSG model (black dashed lines) and from MLT (solid colored lines) when $P_{\rm turb}$ is neglected (i.e. when $P=P_{\rm therm}$ and $\nabla\equiv\nabla_\mathrm{therm}=d\ln T/d\ln P_{\rm therm}$). 
Right: Impact of turbulent pressure on the mixing length calibration.  Comparison of superadiabaticity (middle right) and entropy 
(lower right) are shown, for our averaged 3D RSG2L4.9 model (black dashed lines) and MLT with 
different $\alpha$ (solid colored lines) as a function of temperature. 
The upper right panel shows $P_{\rm turb}/P_{\rm therm}$ (dotted line) and $d\ln P_{\rm therm}/d\ln P_{\rm tot}$ (dash-dot line), which following \citet{Henyey1965} are then included in the MLT calculation and in the averaged 3D $\nabla$. The grey dashed line in the middle right panel gives $d\ln T/d\ln P_{\rm therm}$, matching the black dashed line on the upper left panel.
All values shown are derived from the time-averaged, shell-averaged density and temperature profiles, as well as the time-averaged luminosity at the simulation outer boundary, beyond day 4500 in RSG2L4.9.  Figures adapted from \citet{Goldberg2022a}.} 
\end{figure}

Using the Athena++ setup described above, \citet{Goldberg2022a} produced the first 3D RHD calibration of MLT in the RSG regime. Despite large deviations from MLT near the surface, there exists a large region in the Athena++ RSG envelopes where the convective motions and fluxes do resemble those described by MLT (i.e. where convection carries a majority of the flux, where the radial entropy gradient is zero or negative, and where positive density fluctuations correlate with inwards radial velocities).
Therefore the convective velocities and thermodynamic gradients (in particular $\nabla\equiv d\ln T/d\ln P$ and the entropy) which are recovered self-consistently from the 3D RHD equations can be directly compared against analytic expectations from MLT in order to calibrate $\alpha$. The velocity profiles were found to be flatter than MLT would predict, likely owing to the non-local nature of the coherent convective plumes, with good order-of-magnitude agreement but large scatter exceeding a factor of 10 variation of the mixing length parameter $\alpha$ (Fig.~\ref{fig:alphavels}). Comparing $\nabla$ and the entropy with different $\alpha$, \citet{Goldberg2022a} recovered high convective efficiencies ($\alpha\approx3-4$). This comparison is shown in Fig.~\ref{fig:MLTcal} for calibrations both without (left) and including (right) turbulent pressure via the method suggested by \citet{Henyey1965} but oft-neglected in 1D stellar evolution codes.
This is consistent with a number of estimates of larger-than-solar values of $\alpha$ in the RSG regime, from observed RSG populations on the HR diagram compared to stellar models \citep{Chun2018} and early supernova colors \citep{Dessart2013}.


\subsubsection{Modeling Shock Breakout Emission in Type II-P Supernovae}

Another novel application of the Athena++ framework is the explosion of these large-scale convective 3D models. As the real explosion would occur beneath the inner boundary of the Athena++ simulation domain, in the setup detailed in \citet{Goldberg2022b}, an initially spherical shock is driven through the Athena++ simulation inner boundary, fed by a 1D thermal bomb explosion deep in a MESA stellar model.

\begin{figure}
\centering
\includegraphics[width=1.\textwidth]{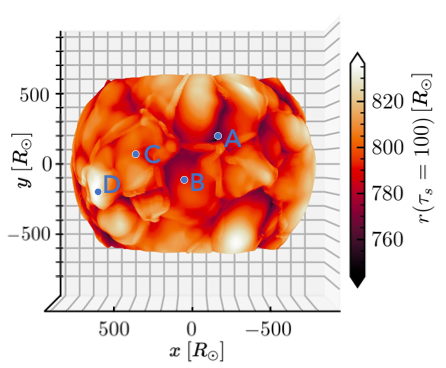} 
\caption{3D surface rendering of the radius where $\tau_s=100$ along each radial direction for the RSG1L4.5 model, where SBO is expected to occur. Four characteristic lines of sight (A,B,C,D) are labeled corresponding to a ``valley", ``hillside", ``plateau", and ``mountain", of the SBO surface.}
\label{fig:SBOSurface}
\end{figure}
In the death of a RSG as a Type II (Hydrogen-rich) Supernova, a strong radiation-dominated shock propagates through the convective envelope. As the shock approaches the outer layers, it reaches optical depth sufficiently low for photons carrying the explosive luminosity to escape ahead of the emerging shock via diffusion, at $\tau\lesssim c/v_{\rm sh}$ where $c$ is the speed of light and $v_{\rm sh}$ is the shock velocity. This leads to a hot ($T>10^5$K), bright flash, known as the ``shock breakout" (SBO). There is an abundance of predictions for the brightness and duration of SBO in spherical-symmetry, (e.g. \citealt{Nakar2010,Rabinak2011,Sapir2011,Katz2012,Sapir2013,Shussman2016b,Sapir2017,Kozyreva2020,Morag2023}). One prediction of the 1D semi-analytics is that the observed SBO duration may be set by the light-travel time over the stellar radius, $R/c\lesssim1$ hour ($\approx$30 min for $800R_\odot$), which if measured would provide a direct constraint on the stellar radius. 
This timescale has not yet been seen in observations, which are currently limited to a few serendipitous detections from the \textit{GALEX} mission, all of which show durations of $\gtrsim6$ hours \citep[e.g.][]{Schawinski2008, Gezari2008, Gezari2010, Gezari2015}. The data are expected to improve significantly with future satellite missions such as \textit{ULTRASAT} \citep{Sagiv2014,Asif2021,Shvartzvald2023}, which is expected to capture hundreds of SBO's at high cadence. 

The 3D envelopes differ from standard 1D models in two important ways.  First is the extended halo of low-density material outside the traditional stellar photosphere, which becomes ionized and contributes non-negligibly to the optical depth as the supernova shock approaches. 
The turbulent pressure in this region extends the scale height of the low-density outer layers discussed in the context of 1D models \citep[e.g.][]{Morag2023}, resembling the exceedingly low densities required to reproduce SBO observations \citep{Schawinski2008}.
The second is the large-scale intrinsically 3D convective granulation, spanning large fractions of the stellar surface and spanning ``valley-to-mountain depths" of $\Delta R\approx 10-20\%$ of the (large) stellar radius in both the CO5BOLD and Athena++ simulations, shown in Figure~\ref{fig:SBOSurface}. 
This global structure is also consistent with tomographic observations as discussed in Section~\ref{sec_cycles}. 

\begin{figure}
\begin{center}
\includegraphics[width=1.\textwidth]{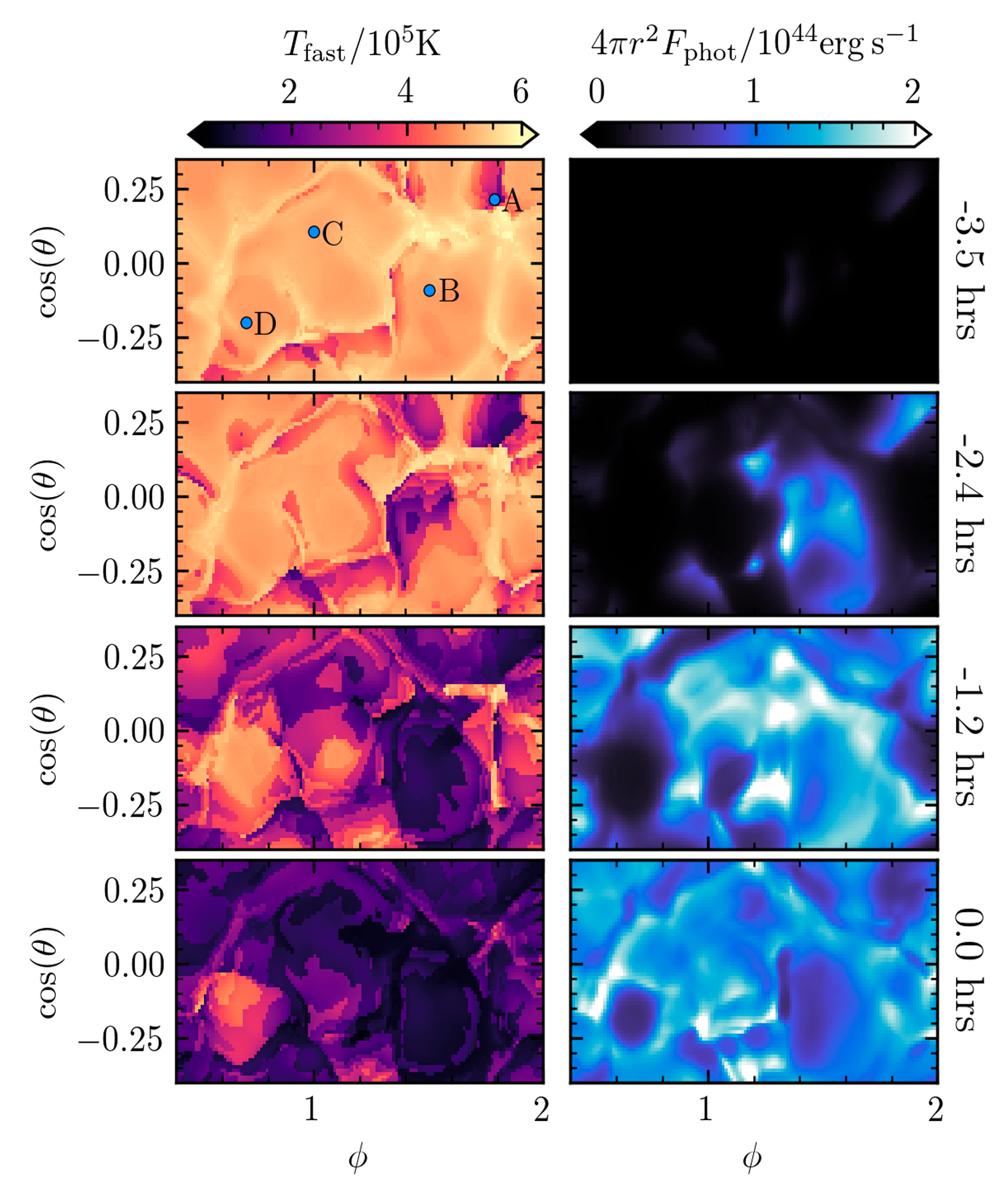}
\caption{Gas temperature of the fastest-moving ejecta along each radial line of sight (left column) and the radiative luminosity $4\pi r^2F_{r}$ at the approximate photospheric location along each radial ray for a given ($\theta, \phi$), defined as the first location where the radiation flux/radiation energy density $F_r/E_r=c/3$ inwards along each radial ray. The temperature of the fastest-moving ejecta signals whether thermal energy is trapped behind the shock pre-``breakout" (indicated by hotter temperatures) or whether it is able to escape and cool (cooler temperatures).
Time until peak bolometric luminosity is shown to the right of each row, and the panels are sampled approximately one diffusion-time apart. Reference locations A, B, C, and D indicated in Fig.~\ref{fig:SBOSurface} are shown in the ($T_\mathrm{fast}$, -3.5 hours) panel. 
Figure adapted from \citet{Goldberg2022b}.}
\label{fig:sbopanels}
\end{center}
\end{figure}


\begin{figure}
\begin{center}
\includegraphics[width=0.9\textwidth]{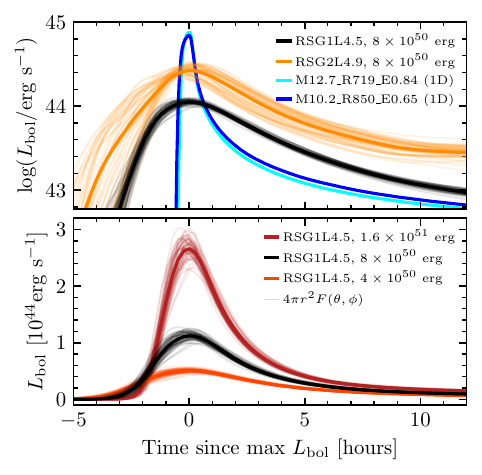}
\caption{Bolometric lightcurves ($L_\mathrm{bol}$) calculated for varied progenitor model (top) and explosion energies (bottom).
Black curves correspond to an $8\times10^{50}$erg explosion of the RSG1L4.5 model (with $\Delta R=80R_\odot$). Faint lines (calculated at $r=2700R_\odot$) show the integrated flux an external observer would see looking back at the star from a particular $\theta,\phi$ location (note that this is not shown at the photosphere as in Fig.~\ref{fig:sbopanels}, but rather for an external observer viewing the whole explosion at different angles from far away). For comparison, SBO lightcurves from spherically symmetric MESA+STELLA explosion models \citep{Goldberg2020} are shown in the upper panel (blues), labeled by ejecta mass/$M_\odot$, radius/$R_\odot$, and explosion energy/$10^{51}\mathrm{erg}$, where the model truncates at the stellar photosphere. The x-axis for all 3D curves is the time since the maximum shell-averaged $L_{\rm bol}$ (i.e. the peak of the thick curves). Figure from \citet{Goldberg2022b}.}
\label{fig:sbolightcurves}
\end{center}
\end{figure}

\citet{Goldberg2022b} find that there are three timescales which can influence SBO. The first timescale is the diffusion time $t_{\rm diff}$ for photons to leak out from the $\tau=c/v_{\rm sh}$ location \citep[e.g.][]{Nakar2010}, which is somewhat prolonged by the presence of the low-density halo above the photosphere. 
The second timescale is $R/c$, which will smear out the SBO signal if larger than $t_{\rm diff}$ \citep[e.g.][]{Katz2012}. The third timescale is introduced by the large-scale 3D granulation which shapes the surface where shock breakout occurs. This can be seen in Figure~\ref{fig:SBOSurface} showing radii of constant $\tau_s=100$ corresponding roughly to the breakout surface $c/v_{\rm sh}$ for shock velocities of a few thousand km/s, where $\tau_s=\int_r^\infty\kappa_s\rho dr$ is the optical depth to electron scattering (with opacity $\kappa_s\approx0.32$ cm$^2$ g$^{-1}$). Four points of varying topography on this surface are labeled A,B,C,D ascending from deepest to shallowest. For the RSG1L4.5 envelope shown, the large-scale inhomogeneity entails radial differences in the SBO surface topography spanning $\Delta R\approx80R_\odot$ (for RSG2L4.9, $\Delta R\approx200R_\odot$). As the shock traverses this large $\Delta R$, it breaks out at different radii at different times along different patches of the simulation domain, 
first in the ``valleys" (e.g. A) and lastly along the ``mountains" (e.g. D).
The third timescale is thus the time for the shock to cross $\Delta R$: $t_{\rm cross}\approx\Delta R/v_{\rm sh}$. 
When $v_{\rm sh}$ is sufficiently large, $v_{\rm sh}>c/(\kappa\rho_0\Delta R)\approx1300$ km/s for $\Delta R\approx80R_\odot$ and a typical breakout density $\rho_0\approx1.25\times10^{-10}\,\mathrm{g\,cm^{-3}}$ and $\kappa\approx0.32\mathrm{cm}^2\,\mathrm{g}^{-1}$, SBO from different patches on the stellar surface at different times will smear out the observed SBO duration. This effect is seen in the evolution of the ejecta temperature and radiative flux at a given patch shown in (Figure~\ref{fig:sbopanels}); ``valleys" (e.g. A) undergo SBO earliest, and even at peak L, some ``mountains" (e.g. D) are still pre-SBO, still hot and comparatively dark. Lightcurves (Figure~\ref{fig:sbolightcurves}) from 3D envelope explosions show a slower rise to peak brightness, with a gradual rise even before half-peak-brightness occurring over multiple hours, with longer durations from half-peak-brightness to half-peak-brightness spanning $\Delta t_{1/2}\approx3-6$ hours. Correspondingly, the predicted luminosities are lowered by a factor of $2-10$ (to $L_\mathrm{bol}\sim10^{44}\mathrm{erg\ s^{-1}}$) as comparable radiation energy escapes over a longer timescale.

Because $t_{\rm cross}$ exceeds $t_{\rm diff}$ except in the limit of low shock velocities and thereby large diffusion times for large $\Delta R$ ($\gtrsim10$\% of the stellar radius in both CO5BOLD and Athena++ simulations), \citet{Goldberg2022b} argue that in either case, $t_{\rm diff}$ and $t_{\rm cross}$ will both exceed $R/c$, rendering it unlikely to measure RSG radii from the SBO duration. However, this introduces the possibility that future observations of SBO can be used to constrain $\Delta R$ and provide yet another probe into the structure and inhomogeneity of the outer layers of cool evolved stars.


\section{Summary}

Convection is a non-local and complex process with non-linear interactions over many disparate length scales happening in a 3-dimensional geometry.
To have a full understanding of this phenomenon, global 3D radiation-hydrodynamics simulations are crucial to modeling the convective atmospheres of RSG and AGB stars. These simulations avoid the use of parameterized convection (e.g., in terms of mixing length theory, commonly used in classic, 1D, linear and non-linear pulsation models) and make it possible to follow the flowing matter, in complete 3D geometry, from the stellar interior to the circumstellar envelope, via the stellar atmosphere and the dust forming region. 

 3D RHD simulations predict how long-lasting giant convection cells (with sizes comparable to the stellar radii), consisting of wide upflow regions that are surrounded by turbulent downdrafts, together with short-lived smaller surface granules and radial fundamental-mode pulsations, give rise to variable surface structures that, as a consequence, affect the triggering and shape of the strong stellar winds as well as their fundamental parameters. These 3D insights can be contextualized and aid in the development and calibration of further 1D models.
 
 From an observational point of view, the dynamical processes across the convective atmospheres largely impact the formation of spectral lines with consequences on photometry, radial velocities, abundances. The associated intrinsic and physical variability can be used to recover the projected velocity field (along the line-of-sight) at different atmospheric layers using the tomographic method and 3D simulations. This theoretical interpretation suggests that the detected hysteresis loops are the characteristic signal of stellar convection and are intimately linked with the acoustic waves originating from the turbulent convective flow propagating upward to the surface layers where the convective energy is modulated by supersonic shocks. \\
 The application of tomography with 3D simulations is a powerful technique for a comprehensive study of the stellar dynamical cycles in evolved stars.

\begin{acknowledgements}
The authors thank the referees for the detailed and constructive comments to improve this work. This work is supported by the French National Research Agency (ANR) funded project PEPPER (ANR-20-CE31-0002). This work was granted access to the HPC resources of Observatoire de la C\^ote  d'Azur $--$ M\'esocentre SIGAMM and resources provided by the
Swedish National Infrastructure for Computing (SNIC) at UPPMAX. 
HPC resources supported by the NASA ATP grant 
ATP-80NSSC18K0560 were provided by the NASA High-End Computing (HEC) program through the NASA Advanced Supercomputing (NAS) Division at Ames Research Center. The Flatiron Institute is supported by the Simons Foundation.

\end{acknowledgements}

\bibliographystyle{spbasic-FS-etal}      
\bibliography{biblio.bib}  


%
%

\end{document}